\documentclass[%
 reprint,
 amsmath,amssymb,
 aps,
nofootinbib, 
]{revtex4-2}

\usepackage{amsmath,amsfonts,bm}

\def\eqref#1{equation~\ref{#1}}

\def\1{\bm{1}}

\DeclareMathAlphabet{\mathsfit}{\encodingdefault}{\sfdefault}{m}{sl}
\SetMathAlphabet{\mathsfit}{bold}{\encodingdefault}{\sfdefault}{bx}{n}

\DeclareMathOperator*{\argmin}{arg\,min}

\usepackage{hyperref}
\usepackage{url}
\usepackage{graphicx}
\usepackage{dcolumn, xcolor}
\usepackage{bm}
\usepackage{subcaption}
\usepackage{float}
\usepackage{makecell}
\usepackage{afterpage}
\usepackage{tabularx}
\usepackage{comment}
\usepackage{algpseudocode}
\usepackage{algorithm}
\usepackage{accents}

\usepackage{balance}

\makeatletter
\DeclareRobustCommand{\cev}[1]{%
  \mathpalette\do@cev{#1}%
}
\newcommand{\do@cev}[2]{%
  \fix@cev{#1}{+}%
  \reflectbox{$\m@th#1\vec{\reflectbox{$\fix@cev{#1}{-}\m@th#1#2\fix@cev{#1}{+}$}}$}%
  \fix@cev{#1}{-}%
}
\newcommand{\fix@cev}[2]{%
  \ifx#1\displaystyle
    \mkern#23mu
  \else
    \ifx#1\textstyle
      \mkern#23mu
    \else
      \ifx#1\scriptstyle
        \mkern#22mu
      \else
        \mkern#22mu
      \fi
    \fi
  \fi
}

\makeatother

\usepackage{array} 
\newcolumntype{C}[1]{>{\centering\arraybackslash}p{#1}}

\usepackage{xcolor}
\usepackage{bbm}
\newif\ifcomments
\commentsfalse
\ifcomments
    \usepackage{ulem}
    \def\twcomment#1{{$\!$\color{magenta} [TW: #1]}}
    \def\twedit#1#2{{\color{red}{\sout{#1}}}{{\color[rgb]{0.0,0.5,1.0}#2}}}
    \def\sacomment#1{{$\!$\color{blue} [SA: #1]}}

    \def\zicomment#1{{$\!$\color{blue} [ZI: #1]}}
    \def\ziedit#1#2{{\color{red}{\sout{#1}}}{{\color{blue}#2}}}
\else 
    \def\twcomment#1{}
    \def\twedit#1#2{#2}
    \def\sacomment#1{}

    \def\zicomment#1{}
    \def\ziedit#1#2{#2}
\fi

\begin{document}


\title{Score-based Diffusion Models for Generating Liquid Argon Time Projection Chamber Images}

\author{Zeviel Imani}
  \email{zeviel.imani@tufts.edu}
\author{Taritree Wongjirad}
  \email{taritree.wongjirad@tufts.edu}
\affiliation{Department of Physics and Astronomy, Tufts University, Medford, Massachusetts\\
The NSF AI Institute for Artificial Intelligence and Fundamental Interactions\\
}

\author{Shuchin Aeron}
\email{shuchin.aeron@tufts.edu}
\affiliation{
Department of Electrical and Computer Engineering, Tufts University, Medford, Massachusetts\\
The NSF AI Institute for Artificial Intelligence and Fundamental Interactions\\
}


\begin{abstract}
For the first time, we show high-fidelity generation of LArTPC-like data using a generative neural network. 
This demonstrates that methods developed for natural images do transfer to LArTPC-produced images, which, in contrast to natural images, are globally sparse but locally dense. 
We present the score-based diffusion method employed. 
We evaluate the fidelity of the generated images using several quality metrics, including modified measures used to evaluate natural images, comparisons between high-dimensional distributions, and comparisons relevant to LArTPC experiments.
\end{abstract}

\maketitle


\section{Introduction}


The Liquid Argon Time Projection Chamber\ziedit{s}{} (LArTPC)~\cite{rubbia1977liquid,chen1976p496} is a particle detector technology utilized in several current and future neutrino experiments~\cite{amerio2004design,anderson2012argoneut,acciarri2017design,acciarri2015proposal,abi2018dune}.
Their wide use in experimental neutrino physics derives from their ability to scale to sizes with length dimensions of tens of meters 
while still being able to resolve the three-dimensional location of charged particle trajectories to the several-millimeter scale.

The active portion of a LArTPC consists of a time-projection chamber (TPC) wherein ionization electrons created by charged particles are drifted toward charge-sensitive devices via an electric field created across the TPC volume.
The charge-sensitive devices include planes of sense wires or readouts capable of recording the 2D locations of ionization~\cite{dwyer2018larpix,asaadi2020first}. 
\twedit{Modeling the processes through which the ionization eventually leads to waveforms in the readout electronics is an important and challenging task~\cite{adams2018ionization}.}{For physics analyses using LArTPCs, the production of Monte Carlo (MC) simulation is a critical challenge. 
One of the primary bottlenecks in creating simulated data is the time required to model the deposition and resulting readout electronics waveforms produced by a given pattern of ionization~\cite{adams2018ionization}.} 
\twedit{One challenge is that the model can have long execution times. For example, simulating an event in the MicroBooNE LArTPC can take upwards of 10-15 minutes per event when including trajectories from cosmic ray particles and particles produced by neutrino interactions.}
{For example, one data acquisition event for an approximately meters-cubed LArTPC requires simulating several millisecond waveforms for 5-10k electronics channels. 
For such an LArTPC operating on the surface, modeling the signals from the energy depositions of O(10) charged particle trajectories requires several minutes to produce.}

\twedit{}{Such MC generation challenges are similar to those faced by particle physics experiments utilizing the Large Hadron Collider (LHC).
  There are now several examples of deep neural network-based algorithms which can balance fidelity to Geant4-produced data and high through-put~\cite{paganini2018calogan,vallecorsa20193d,erdmann2019precise,belayneh2020calorimetry,krause2021caloflow}.
  In addition to serving as faster surrogates, there are also potential uses of generative models for the calculation of likelihoods of low-level data which opens the door to novel approaches to various kinds of inference problems.}
\ziedit{}{
As a result, there has been significant activity in developing generative models for LHC experiments, with Generative Adversarial Networks (GANs)~\cite{10.21468/SciPostPhys.8.4.070,de2018controlling,DERKACH2020161804,alanazi2022machine,rehm2021reduced,buhmann2022hadrons,bieringer2022calomplification,anderlini2023towards,hashemi2023ultra,li2023generative}
and Normalizing Flows
~\cite{lu2021sparse,butter2023generative,diefenbacher2023l2lflows,raine2023nu,golling2023interplay,xu2023generative,krause2023anomaly}
having found a wide range of applications.
However, as far as we are aware, there has not been as much progress on generative models for LArTPC data.
In our experience, many of these approaches struggled to generate images from our LArTPC dataset. 
Our previous work~\cite{lutkus2022towards} using another generative modeling approach, Vecter Quantized-Variational Auto Encoders or VQ-VAEs, did show some promise in beginning to produce images with trajectories that resembled those in LArTPCs. 
But in our judgment, these images still lacked sufficient image quality. 
}
In this paper, we describe results using a generative diffusion model.
We find that it produces images of particle trajectories that are very similar to those from a Geant4-based simulation. 
As far as we \ziedit{}{are} aware, this is the first demonstration of such quality for LArTPC data.

Currently, diffusion models (DMs) are attracting much attention due to their ability to generate novel images with high fidelity and large variance.
These models are being used for a large variety of image types, including natural images~\cite{rombach2022high}, paintings~\cite{yi2021exploring}, or drawings~\cite{peng2023difffacesketch}.
When asked to generate novel images, they produce samples that match closely to the style and subjects of the images from the training set.
They also produce samples that vary over the different semantic content included within the training set.
\ziedit{In HEP, diffusion models}{For example, in high energy physics DMs} have primarily been used in collider physics for jet unfolding ~\cite{leigh2023pc, diefenbacher2023improving, butter2023jet}, and calorimeter data generation ~\cite{vinicius2022Calorimeter, buhmann2023caloclouds, acosta2023comparison}. 
\twedit{Given their versatility, DMs will likely find use in many other applications.}{} 

In our work, we analyze the results of training a \ziedit{explore}{} score-based diffusion model~\cite{song2021scorebased}, 
which is one of several implementation approaches.
We find that this model can generate images which reproduce the features, 
both at high and low distance scales, 
found within LArTPC-like images containing particle trajectories.
\ziedit{Such images are very different from those for which DMs were developed.}{Such images are globally sparse but locally dense, making them very different from the natural images for which DMs were developed.}
Furthermore, the way we measure the quality of the images produced by DMs for LArTPCs will differ from those for natural images.
Therefore, we also present methods of quantification we believe are relevant to
LArTPC images.
As a first proof-of-principle exploration, 
we only train and generate images, which are 64x64 pixels in size. 
In the context of current experiments, this would be considered small as LArTPCs
produce images on the scale of 1000x1000 pixels.
Our \twedit{aim is to investigate if}{aim is to show that} DMs can generate images of high enough quality such that DMs may one day serve as surrogate models for LArTPC detector simulations. 
However, this work does not yet address the desire to improve the speed of generating MC simulation. 
The challenges and potential directions toward that and other goals will be discussed in later sections.

 \begin{figure*}[t!]
    \centering
    \includegraphics[width=0.85\textwidth]{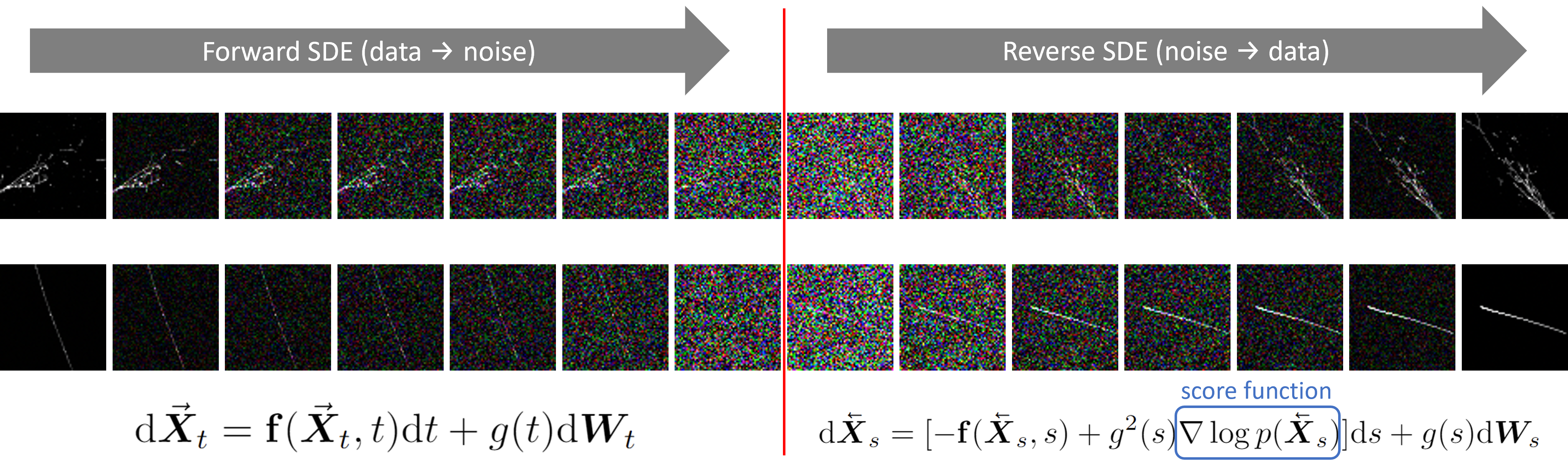}
    \caption{Visualization of the diffusion processes used to train and sample a generative model for LArTPC-like images. The generative process takes noise images (middle columns of the figure) to images that contain particle trajectories similar to what might be seen in a LArTPC. This generative process can be seen as the reverse of the forward process defined by the stochastic differential equation (SDE) on the left, which transforms LArTPC images into noise images. Implementing the reverse process SDE (on the right) from noise to data requires estimating the score function -- a quantity related to the underlying probability distribution of the LArTPC images. The score function is learned through training with data. For simplicity, we have defined a backward-time variable $s = T - t$.
    }
    \label{fig:full_process}
\end{figure*}

\section{Score-based Generative Models}

The goal of generative models is to learn a probability distribution given samples (data) and to produce, i.e., generate new samples from this learned distribution. In most cases, especially when using deep networks, the model learned is an \emph{implicit} one, in that an explicit characterization of the distribution is not sought, only a means to efficiently sample. For a survey of various methods, we refer the reader to \cite{bond2021deep, yang2023diffusion}.

In our case, the data are (portions of) images that come from a LArTPC detector. In this context, we note the following points. 

\begin{itemize}
    \item Compared to the popular image datasets, the LArTPC images are \emph{extremely} sparse, and thus the latent dimensionality of the data manifold is potentially much smaller.
    Thus, the models we construct must be able to produce samples from this low-dimensional manifold where our data resides.
    \item  Furthermore, the frequency at which different semantic content, i.e., different modes of the distribution, appears must mimic that in the reference dataset.
    For example, in the context of LArTPC images, one would want to sample images that depict low-angle scattering by a muon at a much lower rate than scattering examples that reverse the direction of the trajectory as that physical scattering rate is much lower.
    \item We might also expect that the generative model can reproduce the distribution of underlying particle momenta used to prepare the reference data set.
\end{itemize}

Motivated by the recent success of the Score-based Generative Models (SGM) in  \cite{song2021scorebased} that are part of a wider family of so-called diffusion models \cite{yang2023diffusion}, in this paper, we employ them for learning a generative model for the LArTPC data and evaluate its performance. On top of the empirical success, several recent works such as \cite{chen2023improved, chen2022sampling, de2022convergence} have shown that given an accurate score estimator under relatively mild conditions, SGM can provably learn to generate high-dimensional \emph{multi-modal} distributions even when the data is supported on low-dimensional manifold, a scenario that has plagued past efforts in generative modeling which are usually prone to suffering from mode collapse. However, apart from the requirement to learn good score estimators in high dimensions, diffusion models come with additional computational complexity, most notably in generating samples from the learned model \cite{zhang2022fast}. 
Below we begin by providing a brief overview of the SGMs, primarily taken from the seminal paper \cite{song2021scorebased}. 

Most generative models involve learning a transformation or a sequence of transformations, i.e., a process, between easy-to-sample distribution, e.g., the normal distribution and the data distribution.  For diffusion models, one connects the data images to the images sampled from the normal distribution
through a diffusion process where at each time step, an increasing amount of noise is added to the data images. The SGM learns to undo this diffusion process, i.e., reverse the forward noising process in a principled way. This reverse process can then be used to generate samples from the manifold of images implied by the reference data set.

To be precise, following \cite{song2021scorebased} forward diffusion process of noising the images is captured by the following Stochastic Differential Equation (SDE) starting at time $t=0$ with initial condition $\vec{\bm{X}}_0 \sim$ data to time $t= T$:
\begin{equation}
    \mathrm{d}\vec{\bm{X}}_t = \mathbf{f}(\vec{\bm{X}}_t,t)\mathrm{d}t + g(t)\mathrm{d}{\bm{W}}_t.
\end{equation}
Here $\vec{\bm{X}}_t \in \mathbb{R}^d$ is a vector-valued random variable containing the pixel values on the $d$ pixels of the image at time $t$. 
$\mathbf{f}(\vec{\bm{X}}_t,t)$ and $g(t)$ are referred to as the drift and diffusion functions, respectively, of the SDE, and $\vec{\bm{W}}_t$ is a standard Wiener process (aka Brownian motion) independent of $\vec{\bm{X}}_t$.

The key idea behind SGM is based on results that under mild conditions the reverse process defined by $\cev{\bm{X}}_{s} = \vec{\bm{X}}_{T-t}$~\footnote{In our notation, bold-face variables represent vector quantities while the vector arrow indicates the whether the time variable is for the forward or backward process.} is also a diffusion process~\cite{anderson1982reverse, Conforti_2022}.
Figure~\ref{fig:full_process} provides a pictorial illustration of the forward (data-to-noise) and backward (noise-to-data) processes.
The reverse SDE is given by the following equation for $s \in [0, T]$ (which, following \cite{chen2023improved}, we write in more transparent forward way):
\begin{multline}
   \label{eq:gen_sde}
    \mathrm{d}\cev{\bm{X}}_{s} = [ -\mathbf{f}(\cev{\bm{X}}_{s},T - t) +
    g^2(T - t)\nabla \log p(\cev{\bm{X}}_{s}) ]\mathrm{d}s \\
    + g(T -t)\mathrm{d}{\bm{W}}_s,
\end{multline}
where ${\bm{W}}_s$ is a standard Wiener process and $p(\cev{\bm{X}}_s)$ is the probability density function for the (backward) process at time $s \in [0, T]$. Note that $p(\cev{\bm{X}}_s) = p(\vec{\bm{X}}_{T-t})$. 

Given the parameters of the forward diffusion process, namely the drift and diffusion functions, realizing the reverse diffusion requires an additional quantity $\nabla \log p(\vec{\bm{X}}_t)$, referred to as the score function \cite{hyvarinen2005estimation}. While we do not have access to this directly (since we only have access to samples from the data distribution) this can be estimated via a time-dependent score estimator denoted, $\mathbf{s}_{\bm{\theta}}(\vec{\bm{X}}_t,t)$ that is parameterized via a deep network with parameters $\bm{\theta}$ -- hence the name Score-based Generative Model (SGM). In particular, in \cite{song2021scorebased} one solves for:
\begin{multline}
    \label{eq:lossfn}
    \bm{\theta}^* = \argmin_{\bm\theta} \int_{0}^{T} \Bigl\{ \lambda(t) \mathbb{E}_{\vec{\bm{X}}_0} \mathbb{E}_{\vec{\bm{X}}_t|\vec{\bm{X}}_0} \Big[ \\
    \big{|}\big{|} \mathbf{s}_{\bm{\theta}}(\vec{\bm{X}}_t,t) - 
    \nabla \log p(\vec{\bm{X}}_t \, | \, \vec{\bm{X}}_0 ) \big{|}\big{|}^2_2 \Big] \Bigr\} dt.
\end{multline}

The expectation with respect to $\vec{\bm{X}}_0$ in \eqref{eq:lossfn} can be approximated with respect to the empirical distribution of $\vec{\bm{X}}_0$, which is precisely supported on the given dataset of images. For the inner expectation, for specific choices of $f(\bm{x}, t)$ and $g(t)$, one can explicitly write down the \emph{closed} form of $\log p(\vec{\bm{X}}_t|\vec{\bm{X}}_0)$. The outermost expectation in \eqref{eq:lossfn} is with respect to a distribution over the time window $[0, T]$ and $\lambda(t)$ is a positive weight function that can be used to normalize or modulate the relative importance of scores estimated at different times. Further details of the choice of the forward SDE, the choice of $\bm{\theta}$ for the score estimator, as well as time discretization of the forward and reverse SDEs and weighting, is discussed in section \ref{sec:training_SGM}.

\section{Data Employed}

We construct 64x64 crops of particle trajectories from images found in the public PILArNet 2D dataset~\cite{adams2020pilarnet}.
This is a public dataset of LArTPC-like 2D images and 3D voxelized data.
The data is made using \textsc{Geant4} to simulate the transport of one or more particles within a volume of liquid argon.
The location of ionization created by the particles is recorded within the volume.
It is important to note that the 2D images are not made using a full detector simulation
that models signals on a readout wire plane such as those found in current LArTPC experiments.
We are not aware of such a model that is available outside of one of the existing LArTPC collaborations.
Instead, images are made by simply projecting the 3D locations of ionizations deposited within a volume onto the XY, XZ, and YZ planes.
However, a simple model of charge diffusion is applied as the ionization deposits are projected to the planes.
This increases the longitudinal and transverse spatial distribution of the deposits.
We use the XZ projections in our studies.

The particles transported to make the images are chosen at random from
electrons, photons, muons, protons, and charged pions.
The starting location of the particles are chosen at random uniformly in the volume.
The momentum of the particles generated are sampled using a uniform distribution whose bounds are particle dependent. Further details on this can be found in Ref.~\cite{adams2018ionization}.
We do not make any cuts on the data set using particle types or momenta.

From the projected images, 64x64 crops are made.  
The locations of crops are chosen to ensure a minimum number of pixels 
above a certain threshold in order to
avoid empty images.
The pixels are then whitened.
A common bias value is subtracted from all the pixels across the data set and then scaled to keep values between -1 and 1. 
Finally, the values were rescaled between 0 and 255 and converted to png files for convenience. 
We prepared a total of 50,000 images for the training dataset. 
An additional 10,000 images were reserved as a validation sample.
We have made our datasets publicly available on  Zenodo\footnote{\url{https://zenodo.org/record/8300355}} for reproducibility and to encourage comparisons with alternate methods. 


\section{Training the Model}
\label{sec:training_SGM}

The choice of forward SDE: Within the SGM framework, one can make different choices for the form of the drift and diffusion functions. In this work, we used what is called a variance preserving SDE (VPSDE) in \cite{song2021scorebased} whose SDE is
\begin{equation}
\label{eq:vpsde}
    \mathrm{d} \vec{\bm{X}}_t = -\frac{1}{2}\beta(t) \vec{\bm{X}}_t \mathrm{d}t + \sqrt{\beta(t)}\mathrm{d}\vec{\bm{W}}_t.
\end{equation}
The function $\beta(t)$ is a time-varying function used to control the amount of noise added at time $t$.

We again follow~\cite{song2021scorebased} and use a linear function in $t$ that varies between configurable bounds, $\beta_{min}$ and $\beta_{max}$:
\begin{equation}
\label{eq:beta}
    \beta(t) = \frac{\beta_{max}-\beta_{min}}{T}t + \beta_{min}. 
\end{equation}
For our choice of values for $\beta_{max}$ and $\beta_{min}$ (and other parameters),
please refer to Table~\ref{tab:configs}.
The function, $\beta(t)$, through its appearance in the drift term of Eq.~\ref{eq:vpsde}, serves to modulate the rate of decay of the mean expected pixel values to zero.
Given some total time interval, $T$, and sufficiently large enough $\beta(t)$ values, this process transforms an initial sample of $\bm{X}_0 \thicksim p_{\textrm{data}}$, drawn from the distribution of data images, into, $\bm{X}_T \thicksim \mathcal{N}(0,\sigma^2\mathbf{I})$. In other words, the pixels at time $T$ are assumed to have become uncorrelated. 
Here, $\sigma^2$ is the variance of the normal distribution and is assumed to be common between all pixels in the images. 

Training our model requires learning the score $\nabla_{\vec{\bm{X}}_t} \log p(\vec{\bm{X}}_t)$ as a function of time.
We use batch stochastic gradient descent to optimize $\bm{s}_{\bm{\theta}}$ according to Eq.~\ref{eq:lossfn}.
For each training iteration, we sample a batch of $N$ (unperturbed) training images, $\vec{\bm{X}}_0$.
For each $\vec{\bm{X}}_0$ in the batch, a random time is sampled from a uniform distribution over the interval $[0,T]$.
Both $\vec{\bm{X}}_0$ and the time, $t$, are used to sample a training example, $\vec{\bm{X}}_t$, through the following transition kernel:
\begin{equation}
\label{eq:apply_noise}
    p(\vec{\bm{X}}_t | \vec{\bm{X}}_0) = 
    \mathcal{N}( \vec{\bm{X}}_0 e^{-\frac{1}{2}\int^t_0\beta(s)ds}, \bm{I}-\bm{I}e^{-\int^t_0\beta(s)ds}).
\end{equation}
Note that the kernel's specific expression is dictated by the choice of a VPSDE as the diffusion process.
By choosing a large $\beta_{max}$, the kernel is driven more quickly towards $\mathcal{N}(\bm{0},\bm{I})$.

The batch of perturbed images, along with the sampled batch of times, $t$, are passed into the score function estimator, $\bm{s}_{\bm{\theta}}(\vec{\bm{X}}_t, t)$, implemented as a convolution neural network (CNN). 
We follow the work of Song et al.~\cite{song2021scorebased} for the CNN architecture and use their Noise Conditional Score Network (NCSN). 
We chose to keep many of the default configurations used in that work
for modeling the CIFAR-10 dataset~\cite{krizhevsky2009learning}.
The full list of parameters can be found in appendix \ref{appendix:traingen_details}, table \ref{tab:configs} with our modifications highlighted.

The implementation of the loss function we use to train $\bm{s}_{\bm{\theta}}(\vec{\bm{X}}_t, \vec{t})$ follows directly from Eq.~\ref{eq:lossfn} and Eq.~\ref{eq:apply_noise}. 
Because of the marginal probability, $p(\vec{\bm{X}}_t|\vec{\bm{X}}_0)$ is Gaussian,
the gradient of the log-marginal probability has a closed-form.
Therefore, the \emph{empirical} loss calculated for a training batch of images is
\begin{equation}
    \label{eq:loss_in_practice}
    \mathcal{L}(\bm{\theta}) 
    = \frac{1}{N}\sum^{N}_{i}|| \bm{s}_{\bm{\theta}}(\vec{\bm{X}}^{i}_{t^{i}}, t^{i}) 
    - \frac{-(\vec{\bm{X}}^{i}_{t^{i}}-\vec{\bm{\mu}}^{i}_{t^i})}{\vec{\bm{\sigma^2}}^i_{t^i}} ||^2_2,
\end{equation}
where the sum is over a batch with $N$ samples, indexed by $i$, in the training batch. 
Each item in the batch uses a different, $t^i$, sampled from a uniform distribution over the interval $t \in (0,T]$.
The values of $\vec{\bm{\mu}}_t$ and $\vec{\bm{\sigma^2}}_t$ are mean and variance in the Normal distribution defined in Eq.~\ref{eq:apply_noise}.
In the training configuration we used, $\lambda(t)$ from Eq.~\ref{eq:lossfn} is chosen to be simply $1$.

We train the model for a designated number of steps, 
periodically saving checkpoints for later use in generating images. 
We will instead refer to training epochs from here onwards, where one epoch is defined as 
the number of training iterations to process the entire training dataset. 
Since our batch size, $N$, is 128 images and our training dataset is 50,000 images, one epoch is equivalent to 391 training iterations. 
We trained the model for a maximum of 300 epochs. 

\begin{figure}[h]
    \centering
    \includegraphics[width=0.4\textwidth]{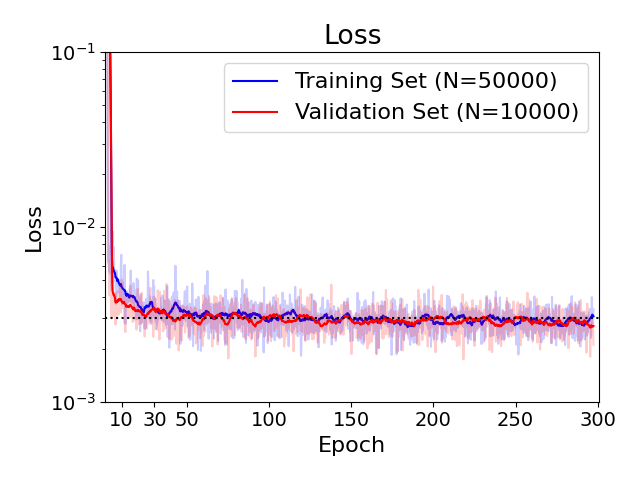}
    \caption{Loss per epoch compared against training set and validation set. The loss function measures the mean-squared error between the predicted and true score functions for a training batch of images. The lighter curve in the background shows the loss for one training batch sampled every 100 iterations. The darker curves in the foreground result from smoothing through a running average over 5 epochs.}
    \label{fig:loss}
\end{figure}

The losses for the training and validation sets versus epoch for the final training run are shown in Figure~\ref{fig:loss}.
The similar loss between the training and validation set as a function of epoch suggests that there is no significant over-fitting of $s_\theta$.
As will be discussed in Sec.~\ref{sec:results}, we observe that the model does not improve appreciably past epoch 50 through several metrics for evaluating the quality of generated images.
The time to train a model for 50 epochs took approximately 6 hours using two NVidia Titian 2080 RTX GPUs.

\section{Generating Images}

After training the model, we were able to produce generated images from any of the pre-designated checkpoints. 
\twedit{Generating 50,000 images from any epoch took approximately 40 hours using two NVidia Titian 280 RTX GPUs.
This method can generate any number of images, we produced 50,000 to match the number of training images for easier comparison.}
{While this method can generate any number of images, we produced 50,000 to match the number of training images for easier comparison.
Generating 50,000 images from any epoch took approximately 40 hours using two NVidia Titian 2080 RTX GPUs. This amounts to about 5.8 seconds per 64x64 image per GPU.}


The generation process, generally, requires implementing a \ziedit{numeral}{numerical} approximation of the reverse diffusion SDE given in Eq.~\ref{eq:gen_sde}. 
We employed a version of the Euler-Maruyama in the code repository of Song et al.~\cite{song2021scorebased}.
Additional approaches were available within the repository. These approaches include alternate numerical solvers and the use of ``corrector" functions, which aim to modify the sampled images at each iteration to be more like the training data.
Exploration of more efficient and accurate SDE methods is still an active area of research for generative models.
We leave studies for optimizing LArTPC image generation for future work.

Generating a sample image under our choice of VPSDE and numerical SDE predictor involves iterating over $M$ constant time steps of $\Delta t=-(T-\epsilon)/(M-1)$ from time $T$ to $\epsilon=1.0\times10^{-3}$.
A non-zero $\epsilon$ is used to avoid numerical instabilities.
During each iteration, $i$, going in reverse order from $i=M$ to $i=1$, 
the sample image, $\vec{\bm{X}}_{t_i}$, 
and predicted mean pixel values $\vec{\bm{\mu}}_{t_i}$ at $t_i=T+(i-M)\Delta t$ are updated using
\begin{equation}
\label{eq:gen_concrete_vpsde_update}
\begin{split}
   \vec{\bm{\mu}}_{t_{i-1}} & =  \vec{\bm{X}}_{t_i} + [-\frac{1}{2}\beta(t_i)\vec{\bm{X}}_{t_i} -\beta(t_i)\bm{s}_{\bm{\theta}}(\vec{\bm{X}}_{t_i},t_i)] \Delta t \\
   \vec{\bm{X}}_{t_{i-1}} & = \vec{\bm{\mu}}_{t_{i-1}} + z\sqrt{-\beta(t_i)\Delta t},
\end{split}
\end{equation}
where $\bm{z}\sim\mathcal{N}(\bm{0},\bm{I})$, and $\beta(t)$ is again defined by Eq.~\ref{eq:beta}.
The pixel values for the prior (noise) image, $\vec{\bm{X}}_{t_M=T}$, are sampled from $\mathcal{N}(\bm{0},\bm{I})$.

We perform a minimal amount of post-processing on the generated images.
By default, the images are normalized to have pixel values ranging from 0 to 255. 
We apply a threshold to each image pixel such that all pixel values below 5 are set to 0. 
This ensures a constant background necessary for the pixel-level sensitivity of our quality metrics, namely the SSNet labels.
A similar threshold is applied to the wire plane waveforms from the LArTPC used in the MicroBooNE experiment~\cite{adams2018ionization}. 
We believe this is an artifact of the diffusion process struggling to achieve the large homogeneous regions of zero values needed for these sparse LArTPC images. 
\ziedit{}{All post-processing is also applied to the training and validation datasets.}

Per our loss curve in Figure \ref{fig:loss}, we can see that the network rapidly improves and then plateaus. 
We have selected epochs 50 and 150 for which we test the quality of generated images. 
After 50 epochs of training, our generated samples are of high fidelity according to our metrics in the results section. 
Extending training to 150 epochs and beyond produces a minimal improvement in the loss and is not worth the extra computational effort. 
As such, we make the case that training for 50 epochs is sufficient. 
Our various quality metrics will further support this choice of training time.

 \begin{figure*}[t!]
    \centering
    \includegraphics[width=\textwidth]{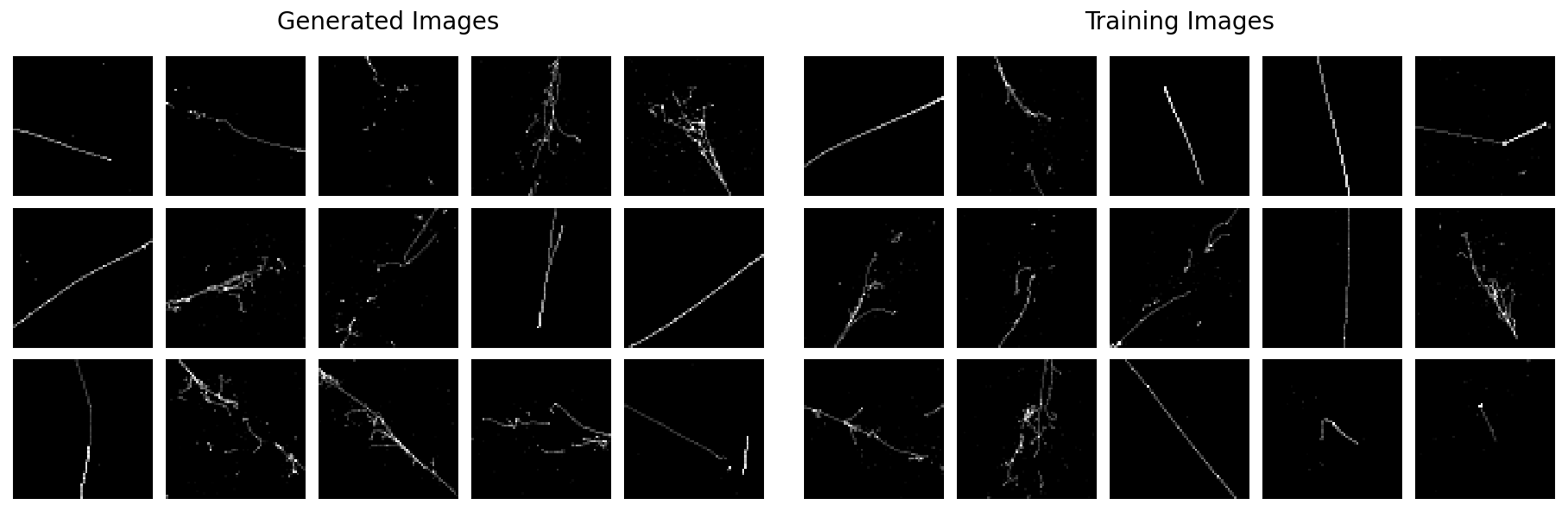}
    \caption{Samples of randomly selected training (left) and generated (right) images. See more samples in Appendix~\ref{appendix:moreimages_mode_check}. 
    }
    \label{fig:compare_training_gen}
\end{figure*}

\section{Evaluating Quality}

\begin{figure}[h!]
    \centering
    \includegraphics[width=0.45\textwidth]{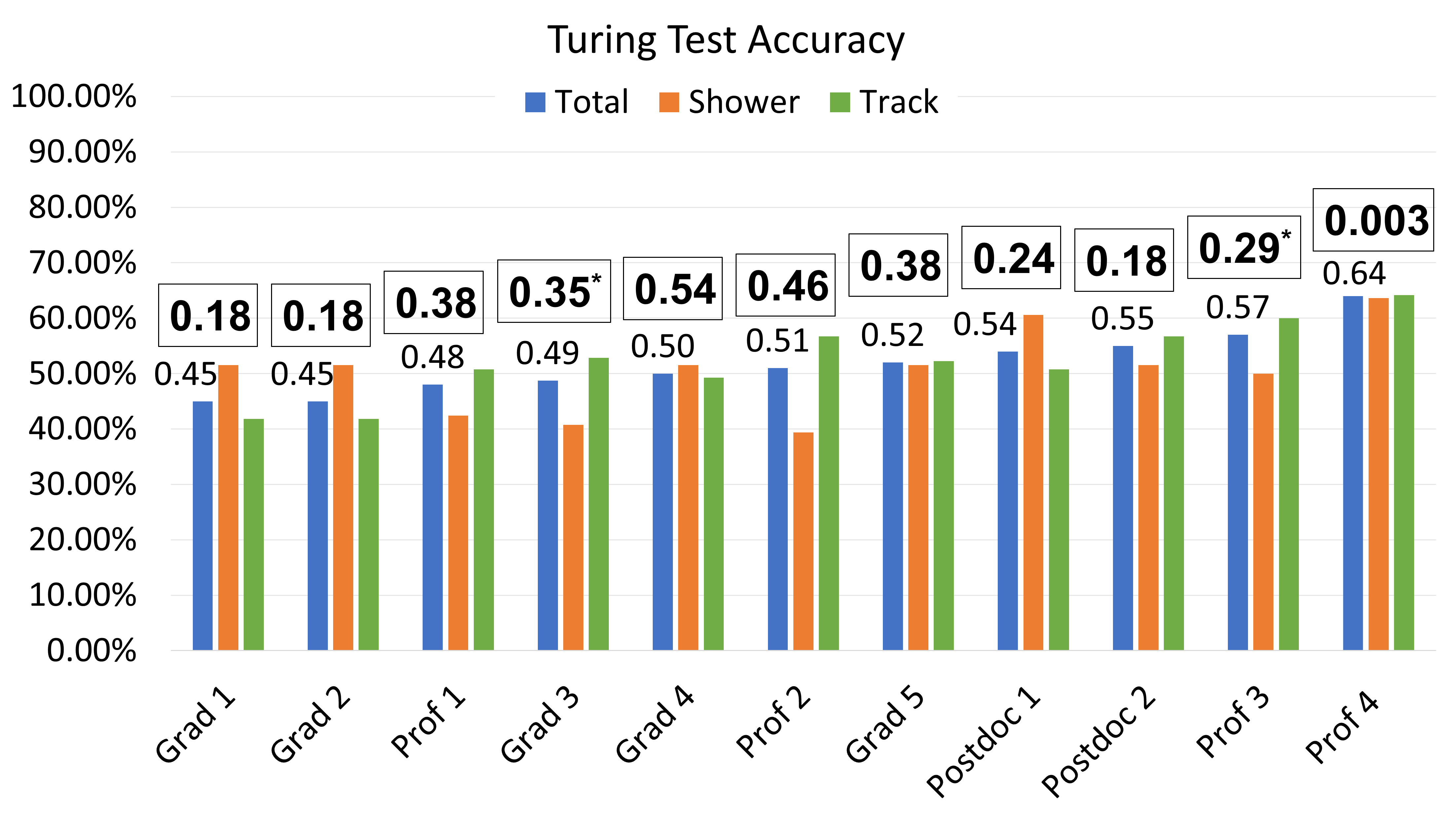}
    \caption{Results from our Turing test where participants were asked to differentiate 50 generated from 50 'true' LArTPC images. 
    The total accuracy is broken down by track (67) and shower (33) images. 
    This corresponds to a binomial distribution where $\protect1 \sigma = \pm 5\%$ from random chance (50\%).  
    P-values for the total accuracy are boxed above each entry. 
    Only one participant was statistically significant, suggesting our generated images are a close match but still have room for improvement.  
    Note that some participants did not complete the full test; their accuracies have been extrapolated, but their p-values (marked with an asterisk) are less significant.}
    \label{fig:Turing}
\end{figure}

We find qualitatively that the images generated by the network are very similar to those in the training set when compared by simple visual inspection. 
In order to quantify this a little further, we conducted a small sample "Turing test" where participants were asked to differentiate true LArTPC (training) images from our generated images by eye. 
We had 11 participants of varying expertise look at 100 randomly selected images generated from epoch 50. 
Only a single participant achieved a statistically significant 64\% accuracy.
Everyone else was within one standard deviation of random chance. 
All the accuracies and corresponding p-values can be found in \ziedit{}{Figure} \ref{fig:Turing}. 
This small-sample experiment suggests that the generated images are at least superficially very similar to the LArTPC images.

While such a qualitative study helps us confirm our belief that the generated model produces high-fidelity images, quantitative measures are desired.
How to measure the quality of images is still an area of open research for natural images.
In addition to how well the images match, we also want to know if the variation in the type of images are similar.
To measure image quality, we have conducted several studies that aim to measure how alike the images are at both large and small scales.

For small scales, we are interested in whether the generated images exhibit the distribution of patterns of pixel values in a small patch.
To measure this, we propose to compare the output of a semantic segmentation network, referred to as SSNet~\cite{collaboration2019deep,abratenko2021semantic}, that has been used by LArTPC experiments as part of their reconstruction and analysis chain~\cite{abratenko2022search}.
The pattern of ionization left behind by charged particles can be classified broadly into two classes: track and shower. "Tracks" are nearly straight lines resulting from more massive charged particles, such as protons, pions, and muons. 
Electrons and photons typically instigate an electromagnetic cascade, resulting in a pattern of deposited charge referred to as a "shower". 
The semantic segmentation network we employ is trained to label individual pixels in an image as either having been created by a particle whose trajectories are track-like or shower-like. 
In order to judge the fidelity of generated images, we compare the distribution of shower and track scores along with the frequency of track-like pixels and shower pixels in the generated and real data set.


For larger scales, we use measures that come from the study of high-dimensional distributions along with reconstructed quantities inspired by physics that are extracted from the images.
For high-dimensional probability distributions, we use the measures: Wasserstein-1 distance, Sinkhorn Divergence, and Maximal Mean Discrepancy (MMD).
These measures attempt to quantify how different the images are at the distribution level.
We were also able to approximate an FID distance metric (details in section \ref{ssec:FID} on the following page).  
For the physics-inspired metrics, we measure quantities that reflect the type of information experiments will want to extract about the particles that ultimately make the images.
This includes a measure of the length and width of track-like trajectories and the energy of shower trajectories.
We use SSNet and simple clustering techniques to identify track-like and shower-like trajectories.
The entire image is then classified as either track or shower based on the majority of pixel labels.
We employ a simple principle-component analysis of the spatial pixel distribution for tracks to measure the length and width. 
Ultimately, the most relevant measure for asking if the data produced by a generative network is sufficient is its impact on physics analyses.
The data we worked with does not lend itself to such analyses and so we leave that to future work to address this.

\begin{figure}[t!]
    \centering
    \includegraphics[width=0.4\textwidth]{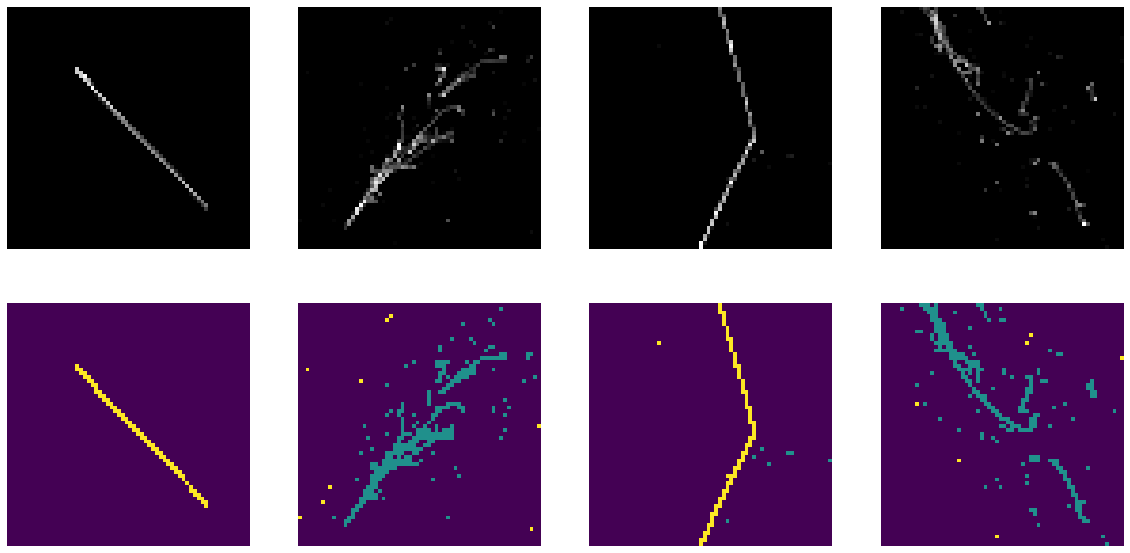}
    \caption{Visualization of the output of the semantic segmentation network (SSNet) used to label track and shower pixels.  The top row shows four LArTPC images from the training set. The bottom row shows the corresponding pixel labels as part of a track-like trajectory (yellow) or shower-like trajectory (cyan).  Track-like trajectories are those produced by muons, charged pions, and protons. Shower-like trajectories are due to electromagnetic cascades initiated by either an electron or the interaction of a photon with the argon. Background pixels where the pixel value falls below some threshold are shown in purple. The output of SSNet on background pixels is not used in our analysis.}
    \label{fig:SSNet_images}
\end{figure}

\section{Results}
\label{sec:results}

The method for determining the quality of LArTPC-like images is still an open question. 
Here, we report on four proposed approaches to quantifying generated LArTPC image quality: 
SSNet pixel-labeling, high dimensional goodness of fit, modified FID score, and physics-based analyses.

\subsection{SSNet Results}

\begin{figure*}[t]
    \centering
    \begin{subfigure}[b]{0.49\textwidth}
        \centering
        \includegraphics[width=\textwidth]{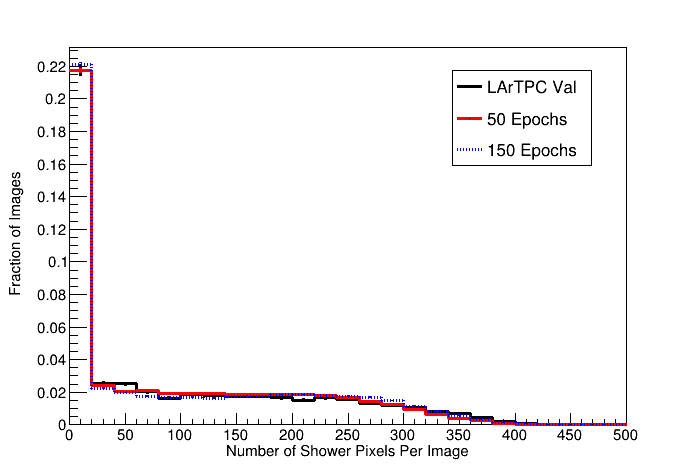}
        \caption{Shower Pixels}
    \end{subfigure}
    \hfill
    \begin{subfigure}[b]{0.49\textwidth}
        \centering
        \includegraphics[width=\textwidth]{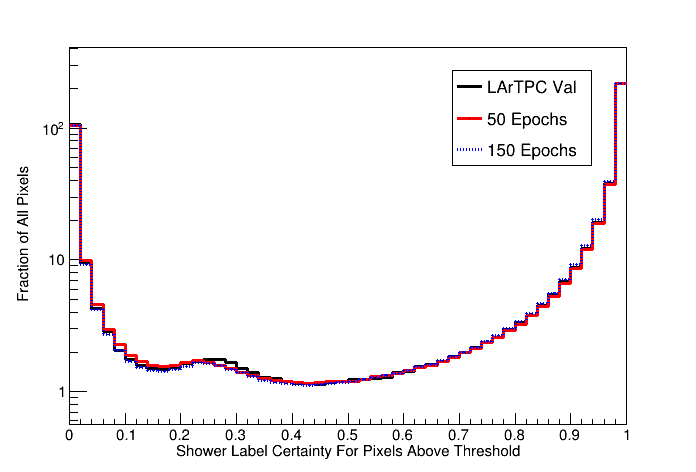}
        \caption{Shower Certainties}
    \end{subfigure}
    \begin{subfigure}[b]{0.49\textwidth}
        \centering
        \includegraphics[width=\textwidth]{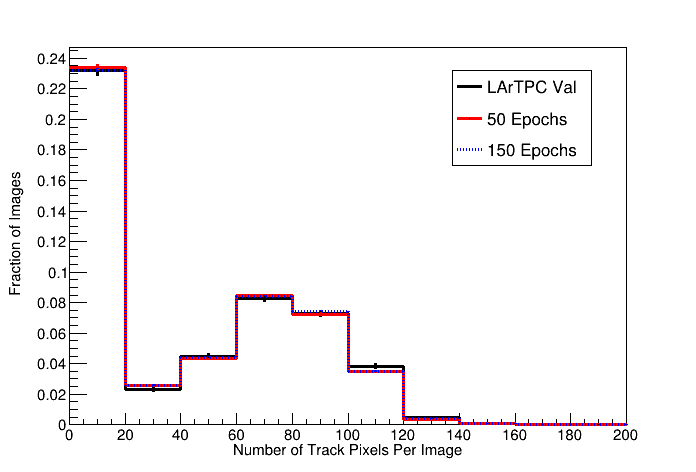}
        \caption{Track Pixels}
    \end{subfigure}
    \hfill
    \begin{subfigure}[b]{0.49\textwidth}
        \centering
        \includegraphics[width=\textwidth]{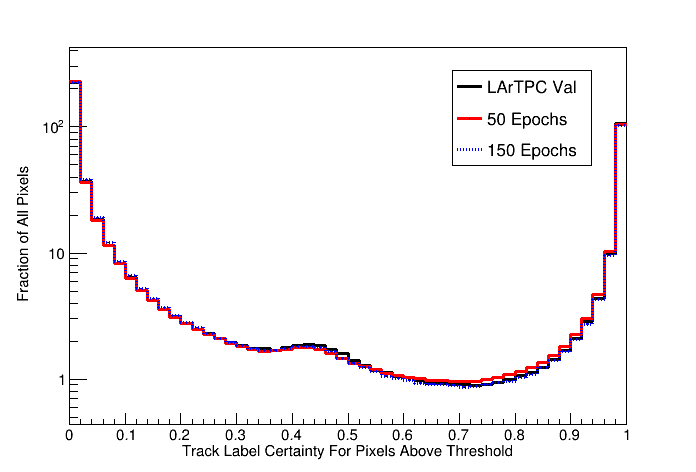}
        \caption{Track Certainties}
    \end{subfigure}
    \caption{SSNet pixel value histograms. (a) shows the number of shower pixels per image, and (b) is the corresponding certainties of these labels. Similarly, (c) is the number of track pixels per image, and (d) is the certainty of the track labels. Across all distributions, we are comparing our generated images at key epochs against our LArTPC validation dataset. We see that \ziedit{epoch 10 is very far from the LArTPC distribution, while}{}epochs 50 and 150 are nearly indistinguishable from each other and the LArTPC validation dataset.}
    \label{fig:SSNet_hists}
\end{figure*}

Our first external metric for gauging the quality of generated images uses our pre-trained semantic segmentation network (SSNet). 
For every image, SSNet labels each pixel as either background, track, or shower, as shown in Figure \ref{fig:SSNet_images}. 
We then compare the distribution of these pixel labels and their associated certainty against our validation dataset.
Per our loss curve (and high dimensional comparison), we have selected two key epochs: 50 and 150 to analyze in figure \ref{fig:SSNet_hists}.
We compare these distributions against our validation dataset. 
The generated images produced using weights from epochs 50 and 150 both have nearly identical distributions to each other and the validation dataset.
This is in agreement with our prediction that there is little, if any, improvement from training beyond 50 epochs. 



\subsection{Model Validation Via Comparing High-dimensional Distributions}

Seeking to refine our model selection we propose to compare the pixel-space distributions, i.e., the distributions in the $64^2$ dimension space, of the training, validation, and generated data via three high-dimensional goodness-of-fit (GoF) measures: Maximum Mean Discrepancy (MMD) \cite{JMLR:v13:gretton12a},  Sinkhorn Divergence (SD) \cite{genevay2018learning, mena2019statistical},  and the Wasserstein-1 distance \cite{Sriperumbudur_2012}. 
Wasserstein distances, also referred to as the optimal transport and earth mover's distance, are based on defining distances based on the minimal cost of coupling via a joint distribution with the given distributions as fixed marginals. 
In some cases, this coupling corresponds to a map, a special type of coupling, which in turn can be interpreted as a transport of mass reshaping one distribution into another. 
MMD distances belong to a family of distances called integral probability metrics, which are defined as a maximum absolute difference, i.e., discrepancy, between the expected value of a class of functions under the two distributions. 
Notably, Wasserstein-1 distance is also an integral probability metric. 
Sinkhorn divergence, while not strictly a distance, interpolates between Wasserstein and MMD distances and is numerically faster to compute as well as has faster rates of statistical estimation \cite{pmlr-v89-genevay19a, mena2019statistical}. 

\begin{figure}[h]
    \centering
    \includegraphics[width=0.45\textwidth]{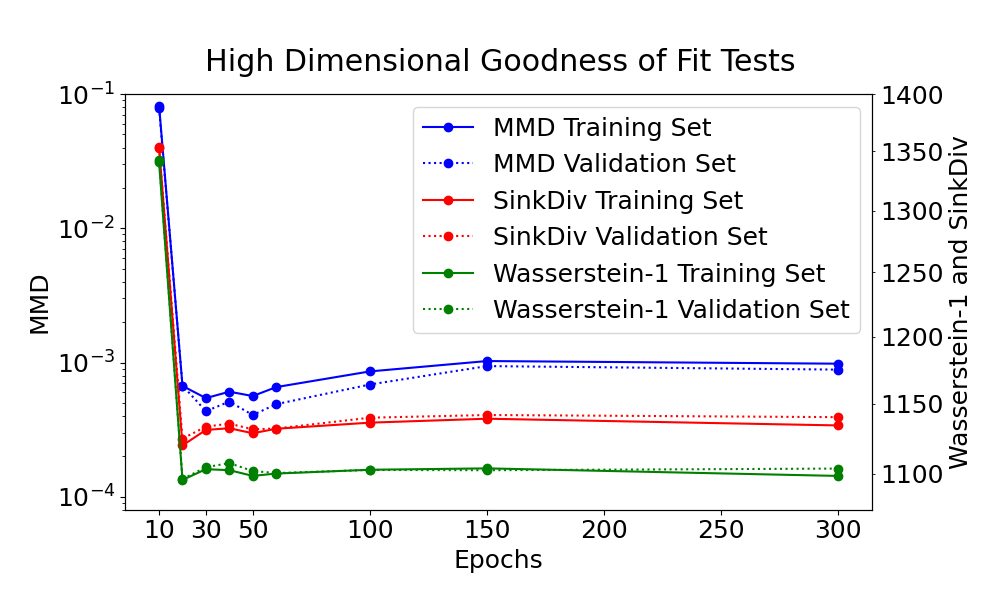}
    \caption{High dimensional goodness-of-fit tests comparing images generated from various epochs against the full training and validation datasets. Explanations of these measures can be found in the text.}
    \label{fig:high_dim_GoF}
\end{figure}

These distribution level comparisons are shown in Figure~\ref{fig:high_dim_GoF}. 
For MMD, the minimum distance is from epoch 50. 
For both SD and Wasserstein-1, epoch 50 is the second minimum after epoch 20. 
The reason for the minimum at 20 epochs is unclear and warrants further exploration in later work. 
This is the case when comparing against both the validation and training datasets. 
We observe that all distance metrics increase slightly with longer training, potentially indicating overfitting. 
However, the variations within this plateau region (starting at 20 epochs) are relatively small compared to the reduction in value from the beginning of training. 
This behavior is similar to the loss curve, supporting our selection of key epochs and reinforcing our choice of epoch 50 as striking the right balance of performance versus training time.



\subsection{Via Inception Score and SSNet} \label{ssec:FID}

\begin{figure}[b]
    \centering
    \includegraphics[width=0.45\textwidth]{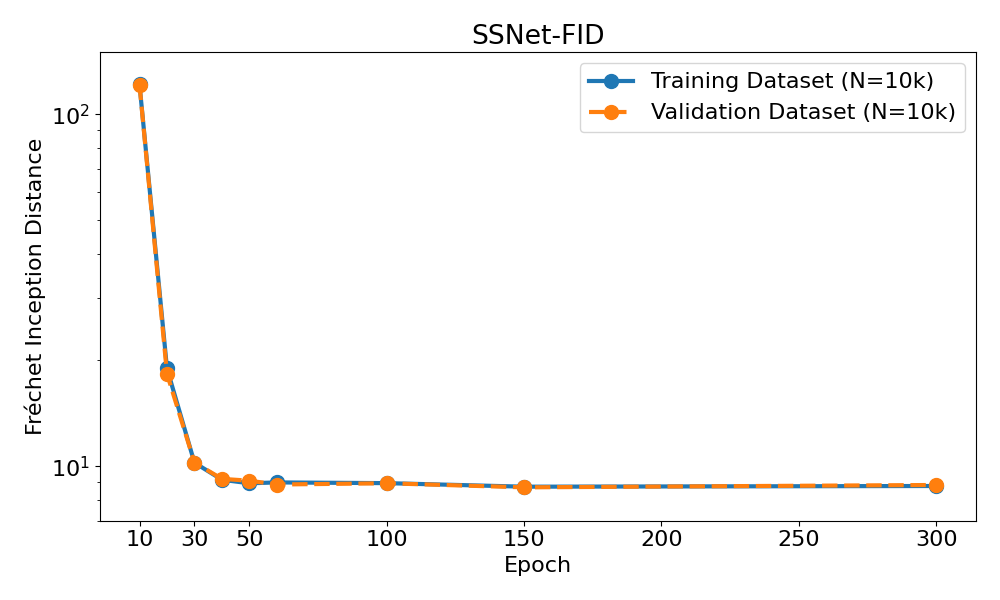} 
    \caption{Our SSNet equivalent of Fréchet inception distance (FID) compares generated images for each epoch against the training set (blue) and validation set (orange). As with previous metrics, we see a rapid improvement and then a plateau in performance. All epochs in the plateau are within a narrow range, with epoch 150 being the minimum. For the convenience of those who might wish to compare future work with our results, the values for each epoch are shown in Appendix \ref{appendix:eval_details}, Table \ref{appendix:SSNet-FID_values}. The minimum is at epoch 150, with all epochs after 50 being very close. 
    \label{fig:FID}}
\end{figure}

A common metric for image generation is the Fréchet inception distance (FID)~\cite{FIDpaper}. 
This is calculated by getting the deepest layer activations of the classifier model for a dataset, fitting the activations to a multivariate Gaussian, and then finding the 2-Wasserstein distance between the training and generated datasets. 
Typically, the activations come from the deepest layer (pool3) of Google's Inception v3~\cite{szegedy2016rethinking}. 
However, we do not have a well-defined classifier for LArTPC type images. 
Instead, we use SSNet and get the activations of one of the deepest convolution layers (double\_resnet[2]) as an 8,192 parameter vector and conduct the prescribed calculations. 
The architecture of SSNet features skip connections, and alternate layer choices quickly become computationally unwieldy. 
Despite this, we believe our SSNet-FID is analogous to traditional FID.
This SSNet-FID metric, shown in Figure \ref{fig:FID}, supports our previous analyses; the model rapidly improves before performance plateaus with an inflection point around epoch 50. 
The best image activations, i.e., closest to the training and validation datasets, are produced by epoch 150, but the differences within the plateau are minimal. 
A table of SSNet-FID values can be found in Appendix~\ref{appendix:eval_details}.

\subsection{Physics-based Comparisons}

An important characteristic of  LArTPC images is the fact that they encode underlying physical processes. 
As such, we have devised a few simple tests to verify that the generated images follow the expected realistic behavior of particle interactions. 
Perhaps the most relevant feature is the energy the particle deposits within the detector. 
We begin by separating the generated images as either track or shower according to the dominant SSNet pixel labeling. 
The different physical processes of track and showers require separate methods of analysis.  

\begin{figure}[h]
    \centering
    \includegraphics[width=0.45\textwidth]{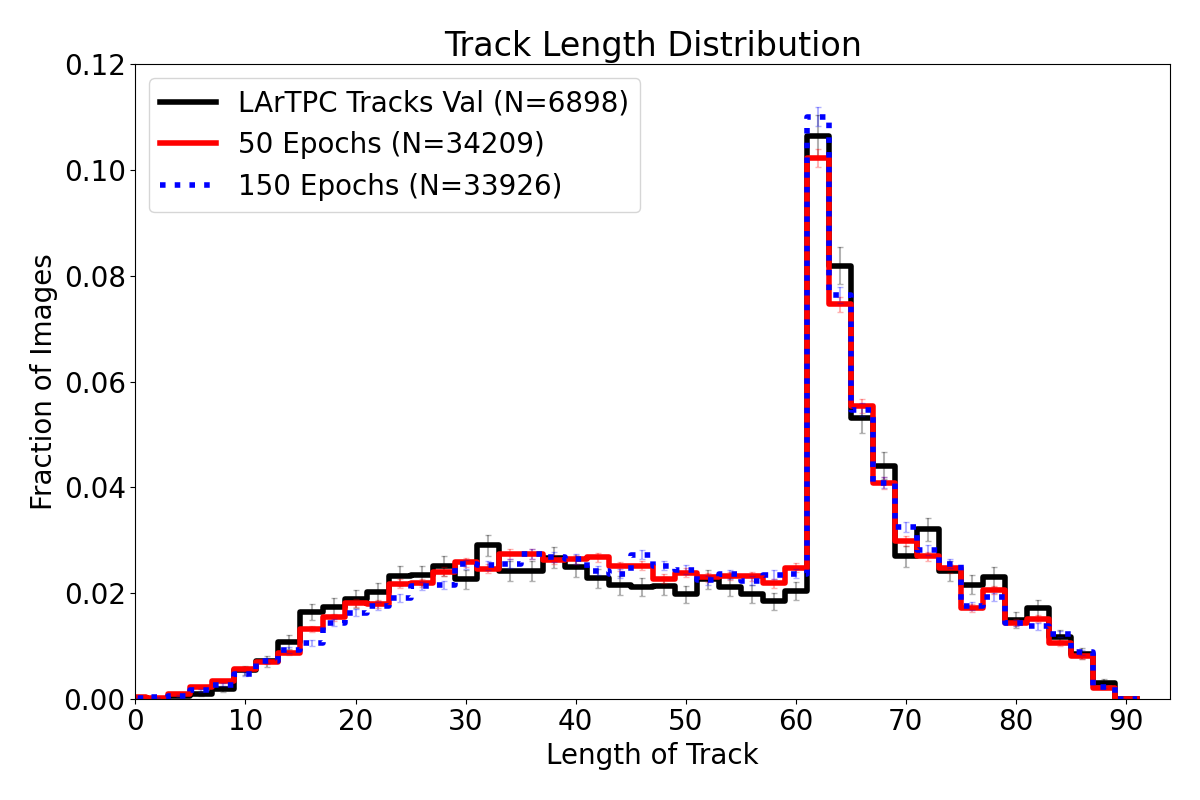}
    \caption{Histogram showing the fraction of track images per length of the track (excluding background noise). The number of track events is shown in the legend. The peak of lengths at 64 results from nearly straight tracks that pass through the entire 64x64 image. 90 is the maximum possible length along the image diagonal.}
    \label{fig:track_length}
\end{figure}

\begin{figure}[h]
    \centering
     \begin{subfigure}{0.21\textwidth}
        \includegraphics[width=\textwidth]{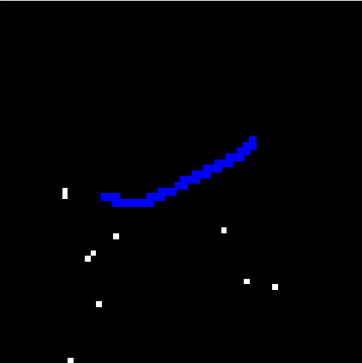}
        \caption{Length 28, Width 5}
    \end{subfigure}
    \hfill
    \begin{subfigure}{0.21\textwidth}
        \includegraphics[width=\textwidth]{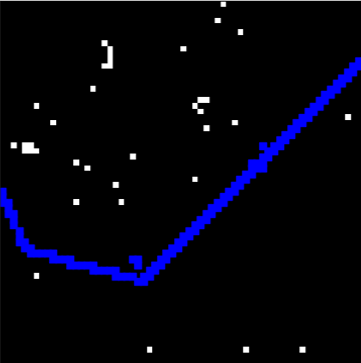}
        \caption{Length 68, Width 27}
    \end{subfigure}  
    \caption{Example tracks from the validation dataset. The track is highlighted in blue, and the \emph{excluded} background noise is in white. Intensities at a maximum for visualization purposes. Track (a) has some minor curvature. Track (b) is composed of three connected tracks and a few pixels of intersecting background noise.}
    \label{fig:example_tracks}
\end{figure}

For track-like events, energy is deposited as it moves through the detector medium (liquid Argon), so we can simply find the length of the track to get an approximate measure of the energy. 
For this calculation, we used DBScan\ziedit{}{~\cite{ester1996density} } (eps=3) to remove background noise and keep only the largest cluster of pixels.
We then use a convex hull algorithm to determine the distance between the two farthest points on the track. 
The majority of track images contain a single particle track.
However, many events have more than one track. 
This is from interaction vertices producing multiple track-producing particles or occasionally overlapping events; an example is shown in Figure \ref{fig:example_tracks}. 
Note that the length of these multiple-track events is the longest straight line distance between two points, \emph{not} a trace of the entire event.  
We apply this calculation across our key epoch generated images and LArTPC validation dataset to provide a consistent metric for comparison. 
We then bin these values into a histogram in Figure \ref{fig:track_length} displaying the fraction of images for each length. 
\ziedit{We can see that epoch 10 is unable to capture the shape of this distribution and vastly overproduces tracks of the maximum length.
This is expected as our training metrics suggested that epoch 10 was insufficiently trained, and the images produced are still noisy.}{}
\ziedit{Epoch 50 and 150 fare much better, and it is difficult to determine which}{It is difficult to determine whether epoch 50 or 150} more closely matches the validation distribution.
To aid our differentiation, we have calculated a total chi-squared comparing each generated histogram to the validation histogram, recording in Table \ref{tab:chi-squared}. 
We find that epoch 50 is a closer match.
The histograms are not a perfect match, but we believe them to be reasonably close enough to justify the functionality of this method of generation. 

We also measured the width of the track by finding the maximum distance along the secondary axis of a principle component analysis applied to the largest cluster of track pixels. 
We are applying the same DBScan clustering algorithm as with the track length analysis. 
We are interested in the track width because it correlates with several physical processes within the detector.
First, in the context of our images, where track trajectories are made by muons, pions, and protons,
the track width is sensitive to the frequency of secondary interactions that can occur with pions and protons, which scatter off nuclei in the detector. In theory, it is also sensitive to the shape and frequency of delta-ray production from muons. The delta rays manifest as shower features in the images and, in principle, should not be included in the track cluster pixels. However, in practice, the delta ray showers are often small enough to be tagged by the SSNet as track pixels and are then seen as track pixel bumps along the track.
Second, the width of tracks as seen in LArTPCs reflects detector physics. This includes diffusion of the ionization cloud as it drifts to the charge-sensitive electronics.  It is also sensitive to the details of how the ionization induces a signal onto the charge-sensitive devices such as wires or pixel-based collection electrodes.

\begin{figure}[h]
    \centering
    \includegraphics[width=0.45\textwidth]{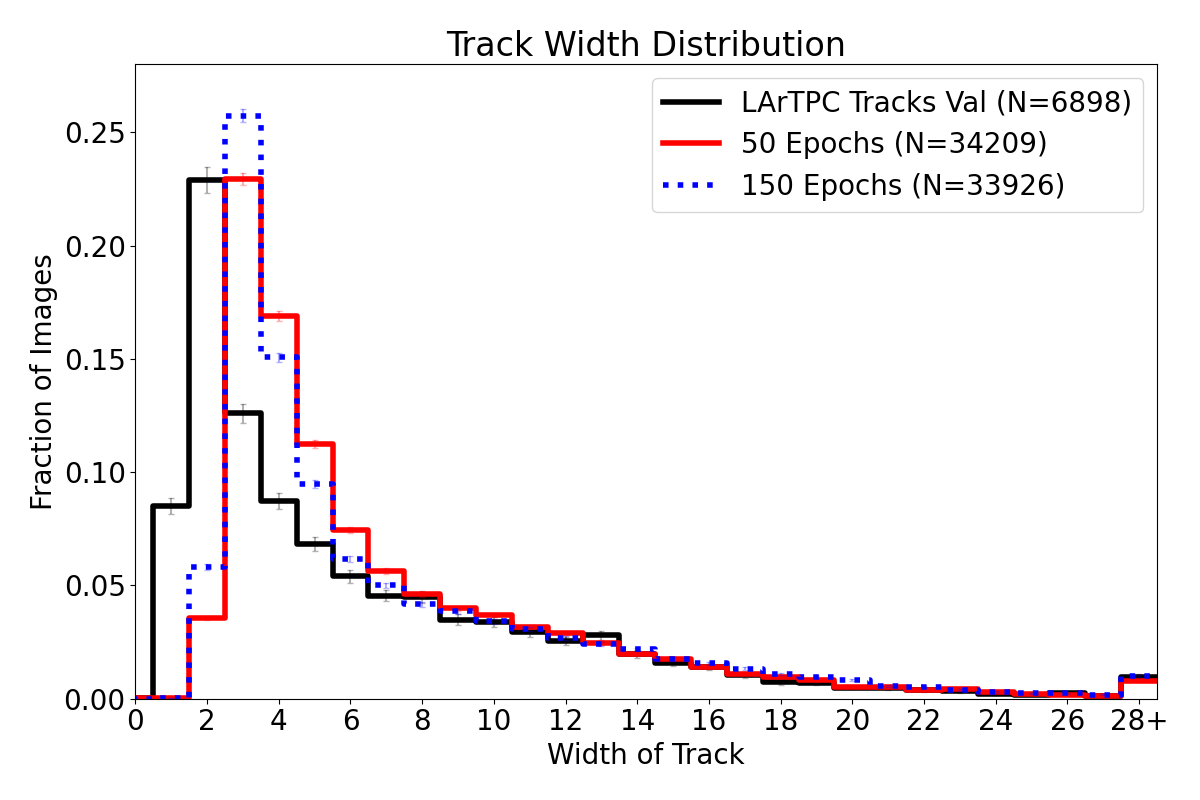}
    \caption{
    The distribution of track widths found in images from the validation dataset and images generated from epochs 50 and 150.  
    Only perfectly vertical or horizontal tracks have a width of 1. 
    The majority of the tracks are 1 - 6 pixels wide. 
    Tracks widths $\ge 28$ have been collected into an overflow bin.
    }
    \label{fig:track_width}
\end{figure}

As before, we have plotted our track width histogram in Figure \ref{fig:track_width}. 
We see that the validation distribution is primarily thin widths (53\% of tracks are $\leq5$ pixels) and then exponentially decays with a long tail.
These thin widths come from straight-line single-track images. 
Wider tracks are caused by track curvature (from processes like multiple-Coulomb scattering) and tracks with secondary interactions. 
All of which can vary slightly if background noise causes bumps or wiggles in the track. 
Figure \ref{fig:example_tracks} shows an example track image from the validation dataset that has multiple tracks and some minor background interference. 
Looking back at our track width histogram, we can compare our key epochs to the validation distribution. 
\ziedit{We find that epoch 10 is unable to reproduce the proper distribution shape and significantly overproduces extremely wide tracks, as evidenced by the final overflow bin. 
A few epoch 10 tracks are so poor quality that the image is classified as entirely noise, in which case the track is reported as having zero width.}{}
The width distributions for epochs 50 and 150 are a \ziedit{closer}{close} fit to the validation dataset.
It is interesting that the large-width tail of the distribution matches fairly well, suggesting that secondary interactions are reproduced fairly well.
However, there are still noticeable differences in the first few bins. 
Disentangling the cause of this difference is reserved for future work where one would want an image generation apparatus capable of better manipulating the activation of physics processes such as delta-ray production, ionization discussion, or the physics of the readout.
Using the chi-squared values in Table \ref{tab:chi-squared}, we find that for the track width distribution agreement, the time epoch 150 is a significantly closer fit to our validation dataset. 
This suggests that longer training might benefit the learning of the more subtle physics that influences this image quality measure.

\begin{figure}[h]
    \centering
    \includegraphics[width=0.45\textwidth]{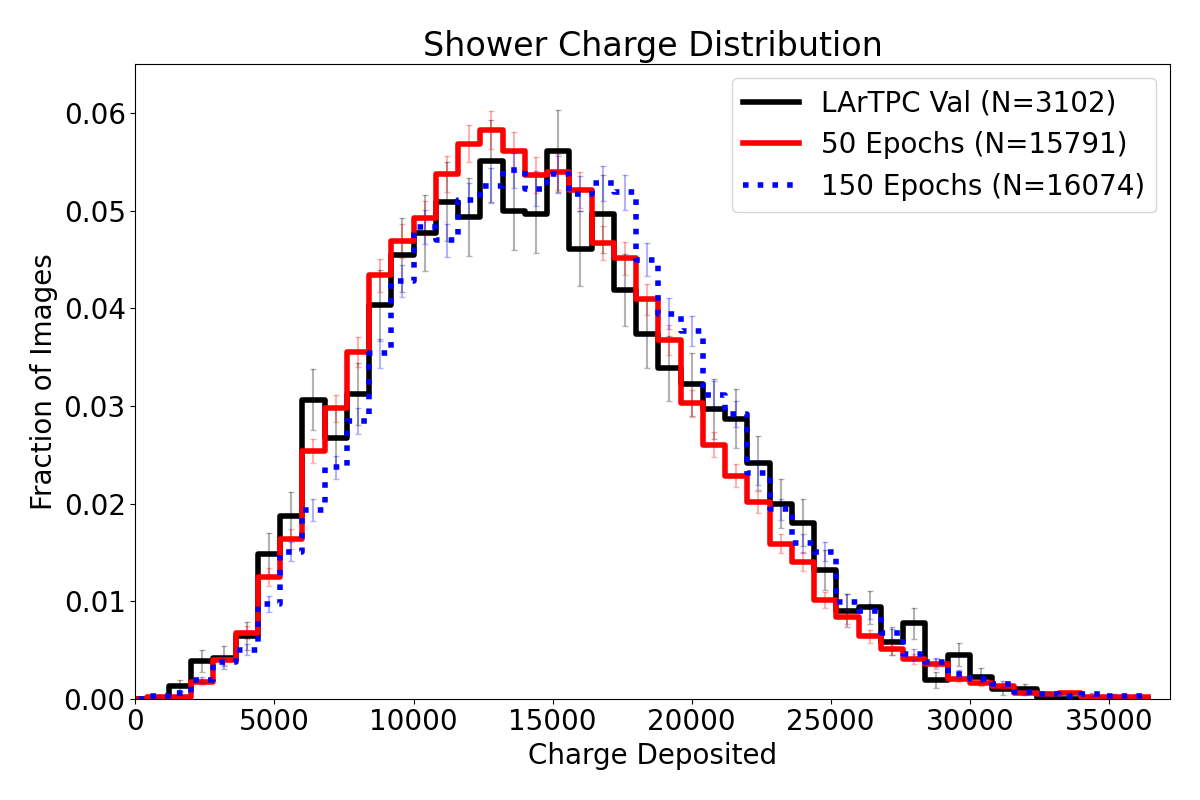}
    \caption{
    \ziedit{A comparison of }{The} total charge deposited in shower event images from epochs 50, 150, and the validation dataset. \ziedit{Three histograms are shown for generated images sampled with models trained over a different number of iterations measured in epochs.}{}
    }
    \label{fig:shower_charge}
\end{figure}

The energy of shower-like events is proportional to the amount of charge they deposit through scattering processes. 
Since our shower images are dominated by these charged particle scattering interactions, we find the energy by summing all the pixel intensities in the image. 
We have plotted a histogram showing the amount of charge deposited per image in Figure \ref{fig:shower_charge}. 
\ziedit{We find that the epoch 10 charge distribution does not resemble the shape of the validation dataset. 
The generated images contain significant background noise, an indication that 10 epochs is insufficient training. 
The surplus in activated pixels correlated to excessively high charge measurements, which we have condensed to a final overflow bin.}{} 
Epochs 50 and 150 both more closely resemble the validation dataset but are not perfect.
Visually, it is difficult to determine which charge distribution is more accurate. 
We can use the mean as a comparison shorthand. 
The shower charge distribution for the LArTPC validation dataset has a mean of 15134, whereas epoch 50 had a mean of 14861 and epoch 150 of 15577. 
This puts epoch 50 as the closer match, off by only 273 compared to epoch 150's difference of 716. 
The mean of a distribution is a simple metric, so we also used a chi-squared test like with the track distributions. 
Per Table \ref{tab:chi-squared},  epoch 50 has a slightly better chi-squared value. 
Both these heuristics agree that epoch 50 produces at least marginally better shower images than our other key epochs. 

\begin{table}[h]
\begin{tabular}{|C{2cm}|C{1.8cm}|C{1.8cm}|C{1.8cm}|}
\hline
$\chi^2$ &
  \begin{tabular}[c]{@{}c@{}} Track \\ Length \end{tabular} &
  \begin{tabular}[c]{@{}c@{}} Track \\ Width \end{tabular} &
  \begin{tabular}[c]{@{}c@{}} Shower \\ Charge \end{tabular} \\ \hline
Epoch 50  & 113   & 399    & 63   \\ \hline
Epoch 150 & 128   & 166    & 70   \\ \hline
\end{tabular}
\caption{Total chi-squared values comparing binned histograms of our generated key epochs against the LArTPC validation set.} 
\label{tab:chi-squared}
\end{table}

Another metric to observe is the makeup of our generated dataset broken down by track and shower images. 
We find that both our training and validation datasets contain 69\% track and 31\% shower images. 
The breakdown of our generated epochs can be inferred from the number of events displayed on the legends in our track length and shower charge plots. 
For convenience, we have calculated the percentages in Table \ref{tab:track_shower_percentages}.
\ziedit{We see that epoch 10 is unable to capture this rate.}{}
Epochs 50 and 150 are within a few percent of the expected rate, but both slightly overproduce shower events.
Images generated from epoch 50 are produced closer to the true ratio of track and shower images than our other key epochs. 

\begin{table}[]
\begin{tabular}{|c|c|c|}
\hline
            & Track \% & Shower \% \\ \hline
Training    & 69.2   & 30.8    \\ \hline
Validation  & 69.0   & 31.0    \\ \hline
Epoch 50    & 68.4   & 31.6    \\ \hline
Epoch 150   & 67.9   & 32.1    \\ \hline
\end{tabular}
\caption{Breakdown of track and shower rates. Images are classified as "track-like" or "shower-like" based on the majority pixel label according to SSNet. The convergence of the track versus shower fraction in the generated image data set produced at the different epochs towards the validation set is a measure of the model's fidelity in reproducing the original data set.}
\label{tab:track_shower_percentages}
\end{table}

\subsection{Checking for Mode Collapse}

\begin{figure*}[t!]
    \centering
    \includegraphics[width=0.7\textwidth]{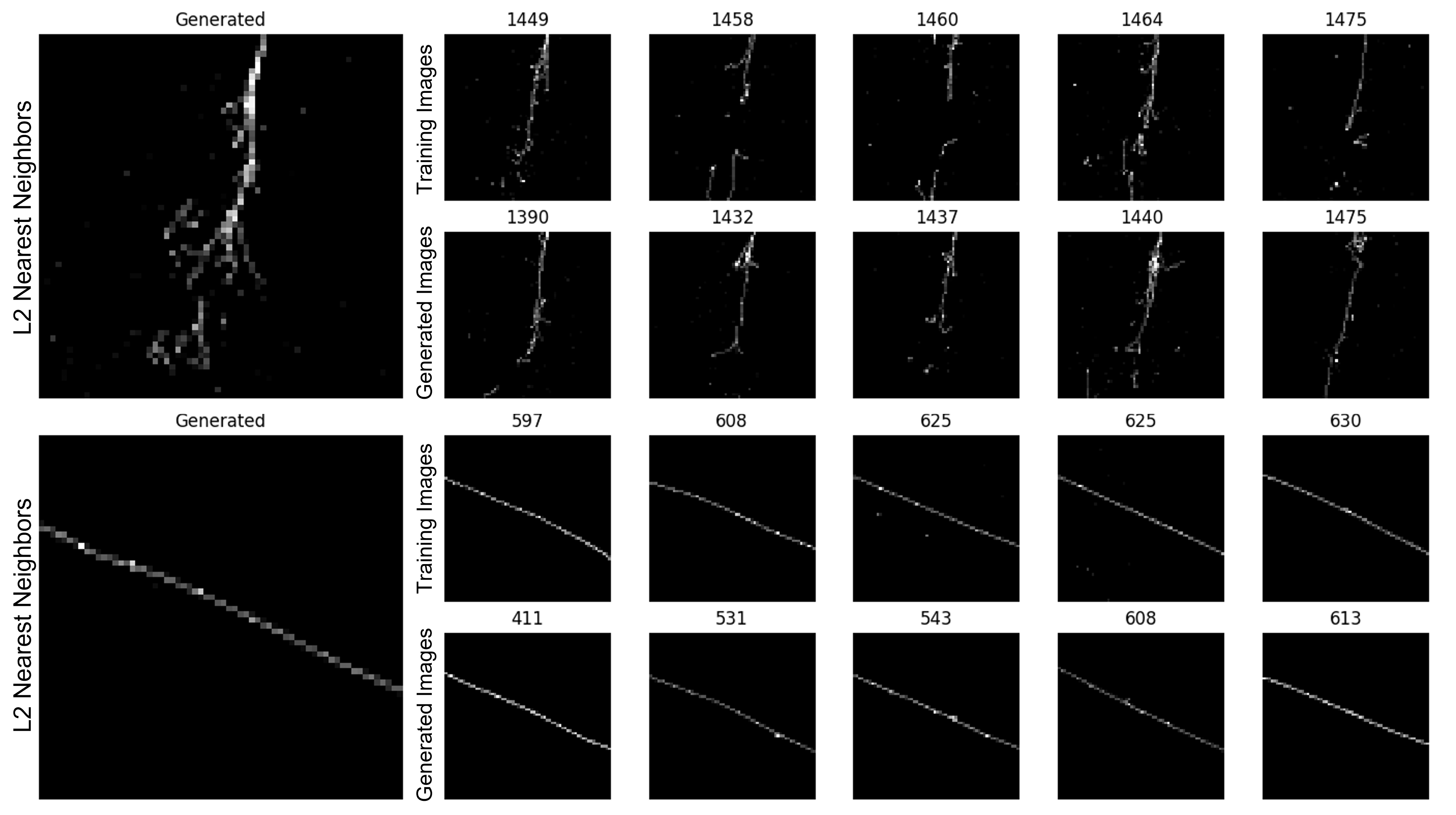}
    \caption{Visual check for mode collapse.  On the left-most column, the large images are an example of a shower-like (top) and track-like (bottom) trajectory generated by our model. For each large image, we use the Euclidean distance (L2 norm) to define the five nearest images in the training dataset (top small row) and generated dataset (bottom small row). The distances are labeled above each comparison image. We have excluded the image itself from the generated dataset. More examples can be found in Appendix~\ref{appendix:moreimages_mode_check}, including using EMD distance.
    \label{fig:l2_two_nearest_neighbors}}
\end{figure*}

A major concern for any method of image generation is mode collapse; the created images need to replicate the characteristics of the training data but still be distinct and unique. 
We want our generated images to sample from the entire space of valid LArTPC images according to the probability distribution $p(\vec{\bm{X}}_0)$ of our training data. 
An example in our context is that we would want the generator to produce tracks smoothly over a range of lengths and angles and not just generate lengths or angles closely around the neighborhood of examples in the training set. 
A useful heuristic is to find the nearest neighbors for a generated image, i.e., the images from the training data that are most similar to a generated image. 
For a single generated image, we calculate its Euclidian distance (L2 norm) from every image in the LArTPC training dataset. 
The (five) images with the smallest distances are defined to be the nearest neighbors. 
This measure of this distance describes a quantitative measure of uniqueness, and qualitatively we can tell that the generated image is unlike the training images. 
We can repeat this procedure to compare a single generated image against the set of all generated images to verify that we are reasonably sampling from the space of images. 
The results of this process can be seen in \ziedit{}{Figure}~\ref{fig:l2_two_nearest_neighbors}.
Several additional examples of nearest neighbors can be found in Appendix~\ref{appendix:moreimages_mode_check}, Figure \ref{fig:l2_neighbors}. 

The use of Euclidian distance when computing distance is standard for typical image datasets such as CIFAR-10 and CelebA. 
However, our dataset of particle trajectories is unlike these standard datasets due to their extreme sparsity and underlying semantic content.
Recent research suggests that Earth (or Energy) Mover's Distance (EMD) is a more meaningful metric for collider-based events \cite{thaler2019physicsemd}. 
As such, we have also found the nearest neighbors using EMD instead of Euclidian distance. 
Examples of this can be found in Appendix~\ref{appendix:moreimages_mode_check}, Figure \ref{fig:emd_neighbors}. 

In both measures of nearest neighbors, we find minimum distances significantly greater than zero, implying that the generated image is unique. 
Visually, we can tell that the nearest neighbors are different images. 
This is particularly evident in shower-like events where the overall orientation of the shower and its density are similar, but the patterns within the shower envelope vary noticeably.
For tracks, the trajectories are very similar, but we do see variation in the frequency and placement of high energy deposits and delta ray features.
Interestingly, the measured distances between the training images and the generated images are similar.
The distance between showers is also larger than with tracks, as one might expect, given the higher inherent variation in the shower process.
One avenue of research is to understand this metric in more detail to both better understand how to quantify potential mode collapse and also for understanding particle trajectories in LArTPCs.
In summary, we believe this qualitative analysis supports the conclusion that the generated images are not mere reproductions of the training images.
Combining this evidence with the reproduction of various distributions, we conclude that the model does not exhibit significant mode collapse, but is instead generating examples with variations similar to that of the training sample.

\subsection{Comparison to Another Model}

As far as we know, the only past attempt at generative models for LArTPCs was 
our previous attempts using Vector-Quantized Variational Autoencoder (VQ-VAE).
At the time, the generated images from this approach began to resemble the patterns seen in LArTPCs, a first compared to past unpublished attempts with GANs and OT approaches. 
However, despite the progress, the generated images were visibly different than the training dataset. 
The study using VQ-VAEs was presented in~\cite{lutkus2022towards} and introduced the use of SSNet outputs to quantify the similarity of generated images to the training sample.
The SSNet measures aligned with observations of a model approaching LArTPC features but not yet closely reproducing them. 
In order to facilitate a comparison to the performance of our diffusion model, we reproduce the SSNet measures and measure the quality of the VQ-VAE images using the approaches presented above.
The full results of this analysis can be found in Appendix \ref{appendix:VQ-VAE}.

\subsection{Associated Code and Dataset Releases}

In order to support efforts to reproduce and compare against our efforts, 
we will prepare a repository of our code, the weights of our generative model, 
and copies of our training, validation, and generated data sets. 
We provide the weights for the final training run at 50 epochs. 
These materials will be stored on Zenodo~\footnote{\url{https://zenodo.org/record/8300355}}.
Scripts and code for training, generating, and calculating our evaluation metrics will be provided on GitHub~\footnote{\url{https://github.com/NuTufts/Score_Based_Diffusion_for_LArTPC_Images}}.

\section{Discussion}

How best to determine the quality of LArTPC-like images remains an open question -- just as is the case for natural images. 
We have proposed several approaches which aim at different views of how to compare the images.
This includes the direct comparison of the images using statistical techniques developed for comparing high-dimensional probability distributions, feature analysis through a domain-relevant CNN (SSNet),
and through the physical quantities extracted from the images.
We find that the VPSDE diffusion model implemented here produces generated images very similar to the validation set.
These measures also largely improve over training time.

One limitation of our quality measures is that they still do not address what we ultimately want to know: would an analysis that attempted to make a physical statement be significantly biased by the use of data created by a generative diffusion model?
Our ability to answer this question is limited by our choice of dataset, which used cropped images that did not make an attempt to capture a full particle trajectory.
This reflects the goal of this work, which was to explore if diffusion models could mimic the collection of features found in LArTPC images.
This study provides encouraging evidence to this narrower question, opening the path towards developing generative models that can potentially assist physics analyses in various ways, thus making the larger question now relevant.

Future directions for our work would aim at addressing the impact on physics analyses but will also need to explore the large space of choices in the implementation of diffusion models.
One direction would be to implement diffusion models for various particle types conditioned on momentum.
Implementation of these would make it possible to generate a collection of images that could be translated and combined to form images representing the final state particles from neutrino interactions.
Particle-level or interaction-level images could then be passed into existing reconstruction and selection algorithms to quantify the amount of bias introduced into those analyses.
An additional technical challenge somewhat unique to LArTPCs is the need to correlate the generation of several objects. 
The most common form of LArTPC in operation uses three sets of wire planes in its readout system.
To fully simulate this, three tomographic projections of a 3D particle trajectory must be generated.
Also, LArTPCs capture not only the ionization left behind by particles but also scintillation light captured in optical sensors, making the problem multi-modal.

Another key direction for development is to improve generation time.
This will be critical if one aims for these models to assist in MC production.
Currently, a full event image from the MicroBooNE experiment used in physics analyses has a size of 1008x3456.
If we assume a linear scaling in complexity from the 64x64 images in this work, 
a naive implementation of our current approach would be computationally infeasible, 
requiring over 68 thousand hours per 2080 RTX GPU to generate 50,000 large images or 81 minutes per image per GPU.  
This time is likely an upper bound, however, as one would imagine more sophisticated procedures such as generating smaller sub-images per particle trajectory and combining them to form a whole event image.
Furthermore, we plan to explore methods used for natural images, 
such as latent space diffusion~\cite{rombach2022high}, which has 
 increased the efficiency for generating higher-resolution images. 
There are possible speed improvements by finding a more optimal implementation of the reverse SDE process.
For example, one can reduce overall generation time by taking larger time steps while integrating the reverse SDE. 
In this strategy, auxillary networks can help correct errors in the state that might reduce the quality of the images~\cite{song2021scorebased}.
Other work centers around the choice of drift and diffusion functions, $\bm{f}$ and $\bm{g}$ \cite{zheng2023fast}.
Another area of active research is into the theoretical connections of diffusion models to optimal transport~\cite{khrulkov2022understanding} and Schrodinger Bridge formalisms~\cite{de2021diffusion},
which may one day guide the choice of drift and diffusion functions for specific goals of the model.
Interestingly, recent work has explored the possibility of extending generative models past the use of SDEs describing diffusion and instead using SDEs that implement other physical processes~\cite{liu2023genphys}.



\ziedit{}{In future work, we plan to expand this score-based diffusion model to larger 512x512 sized images and condition the generated image on particle type and momentum. 
This will entail a significant computational challenge, however, we believe the success in large-scale natural image generation will be reproducible for our LArTPC data. 
These additions might aid existing physics analysis by providing an on-the-fly event generator. 
In the near term, there are potential applications for these models 
in tasks like generating additional background events 
or boosting the performance of current reconstruction algorithms by better conditioning low-level data by removing noise or filling detector gaps. 
We will continue to explore the applicability of these score-based DMs in the space of neutrino detector experiments.  
}


\section{Conclusions}

In this work, we show that a choice of implementation for a score-matching diffusion model is able to learn and generate novel images containing features that would be observed in data from LArTPCs.
We have proposed metrics for future work to determine the model selection and to quantify how well the generated images match the original data set. 
According to these metrics, this is the first example of LArTPC experiments where images produced by a generative model are shown to be very similar to the training set. 
The slight but significant deviation in the generated images is seen in the distribution of narrow track widths which indicates further work to fully capture the physical processes that influence that aspect of the trajectories.
\twedit{}{The above focuses on the quality aspects of generated images and leaves developments in generation efficiency to future work.}
However, we believe our results are an important proof-of-principle demonstration of quality that establishes the groundwork to further pursue the use of score-matching diffusion models for LArTPC experiments.

\subsubsection*{Acknowledgments} 



This material is based upon work supported 
by the U.S. Department of Energy (DOE)
and the National Science Foundation (NSF).
T.W. was supported by the U.S. DOE, 
Office of High Energy Physics under Grant No. DE-SC0007866. 
S.A. was funded by the NSF under CAREER Award No. CCF:1553075.
This work was also supported by the National Science Foundation under Cooperative Agreement PHY-2019786 (The NSF AI Institute for Artificial Intelligence and Fundamental Interactions, \href{http://iaifi.org}{http://iaifi.org}).

\bibliography{mybib_v2}

\begin{thebibliography}{70}
\providecommand{\natexlab}[1]{#1}
\providecommand{\url}[1]{\texttt{#1}}
\expandafter\ifx\csname urlstyle\endcsname\relax
  \providecommand{\doi}[1]{doi: #1}\else
  \providecommand{\doi}{doi: \begingroup \urlstyle{rm}\Url}\fi

\bibitem[Rubbia(1977)]{rubbia1977liquid}
Carlo Rubbia.
\newblock {The Liquid-Argon Time Projection Chamber: a New Concept for Neutrino Detectors}.
\newblock Technical report, CERN, Geneva, 1977.
\newblock URL \url{https://cds.cern.ch/record/117852}.

\bibitem[Chen et~al.(1976)Chen, Condon, Barish, and Sciulli]{chen1976p496}
H.~H. Chen, P.~E. Condon, B.~C. Barish, and F.~J. Sciulli.
\newblock {A Neutrino Detector Sensitive to Rare Processes}: {I. A Study of Neutrino Electron Reactions}.
\newblock 5 1976.

\bibitem[Amerio et~al.(2004)Amerio, Amoruso, Antonello, Aprili, Armenante, Arneodo, Badertscher, Baiboussinov, Ceolin, Battistoni, et~al.]{amerio2004design}
S~Amerio, S~Amoruso, M~Antonello, P~Aprili, M~Armenante, F~Arneodo, A~Badertscher, B~Baiboussinov, M~Baldo Ceolin, G~Battistoni, et~al.
\newblock Design, construction and tests of the {ICARUS} {T600} detector.
\newblock \emph{Nuclear Instruments and Methods in Physics Research Section A: Accelerators, Spectrometers, Detectors and Associated Equipment}, 527\penalty0 (3):\penalty0 329--410, 2004.

\bibitem[Anderson et~al.(2012)Anderson, Antonello, Baller, Bolton, Bromberg, Cavanna, Church, Edmunds, Ereditato, Farooq, et~al.]{anderson2012argoneut}
C~Anderson, M~Antonello, B~Baller, T~Bolton, C~Bromberg, F~Cavanna, E~Church, D~Edmunds, Antonio Ereditato, S~Farooq, et~al.
\newblock The {A}rgo{N}eu{T} detector in the {NuMI} low-energy beam line at fermilab.
\newblock \emph{Journal of Instrumentation}, 7\penalty0 (10):\penalty0 P10019, 2012.

\bibitem[Acciarri et~al.(2017)Acciarri, Adams, An, Aparicio, Aponte, Asaadi, Auger, Ayoub, Bagby, Baller, et~al.]{acciarri2017design}
R~Acciarri, C~Adams, R~An, A~Aparicio, S~Aponte, J~Asaadi, M~Auger, N~Ayoub, L~Bagby, B~Baller, et~al.
\newblock Design and construction of the {MicroBooNE} detector.
\newblock \emph{Journal of Instrumentation}, 12\penalty0 (02):\penalty0 P02017, 2017.

\bibitem[Acciarri et~al.(2015)Acciarri, Adams, An, Andreopoulos, Ankowski, Antonello, Asaadi, Badgett, Bagby, Baibussinov, et~al.]{acciarri2015proposal}
R~Acciarri, C~Adams, R~An, C~Andreopoulos, AM~Ankowski, M~Antonello, J~Asaadi, W~Badgett, L~Bagby, B~Baibussinov, et~al.
\newblock A proposal for a three detector short-baseline neutrino oscillation program in the fermilab booster neutrino beam.
\newblock \emph{arXiv:1503.01520}, 2015.

\bibitem[Abi et~al.(2018)Abi, Acciarri, Acero, Adamowski, Adams, Adams, Adamson, Adinolfi, Ahmad, Albright, et~al.]{abi2018dune}
B~Abi, R~Acciarri, MA~Acero, M~Adamowski, C~Adams, D~Adams, P~Adamson, M~Adinolfi, Z~Ahmad, CH~Albright, et~al.
\newblock The {DUNE} far detector interim design report volume 1: Physics, technology and strategies.
\newblock \emph{arXiv:1807.10334}, 2018.

\bibitem[Dwyer et~al.(2018)Dwyer, Garcia-Sciveres, Gnani, Grace, Kohn, Kramer, Krieger, Lin, Luk, Madigan, et~al.]{dwyer2018larpix}
Daniel~A Dwyer, M~Garcia-Sciveres, D~Gnani, C~Grace, S~Kohn, M~Kramer, A~Krieger, CJ~Lin, Kam~Biu Luk, P~Madigan, et~al.
\newblock Larpix: Demonstration of low-power 3d pixelated charge readout for liquid argon time projection chambers.
\newblock \emph{Journal of Instrumentation}, 13\penalty0 (10):\penalty0 P10007, 2018.

\bibitem[Asaadi et~al.(2020)Asaadi, Auger, Ereditato, Goeldi, Kose, Kreslo, Lorca, Luethi, Rudolf Von~Rohr, Sinclair, et~al.]{asaadi2020first}
Jonathan Asaadi, Martin Auger, Antonio Ereditato, Damian Goeldi, Umut Kose, Igor Kreslo, David Lorca, Matthias Luethi, Christoph Benjamin~Urs Rudolf Von~Rohr, James Sinclair, et~al.
\newblock First demonstration of a pixelated charge readout for single-phase liquid argon time projection chambers.
\newblock \emph{Instruments}, 4\penalty0 (1):\penalty0 9, 2020.

\bibitem[Adams et~al.(2018)Adams, An, Anthony, Asaadi, Auger, Bagby, Balasubramanian, Baller, Barnes, Barr, et~al.]{adams2018ionization}
C~Adams, R~An, J~Anthony, J~Asaadi, M~Auger, L~Bagby, S~Balasubramanian, B~Baller, C~Barnes, G~Barr, et~al.
\newblock Ionization electron signal processing in single phase lartpcs. part i. algorithm description and quantitative evaluation with microboone simulation.
\newblock \emph{Journal of Instrumentation}, 13\penalty0 (07):\penalty0 P07006, 2018.

\bibitem[Paganini et~al.(2018)Paganini, de~Oliveira, and Nachman]{paganini2018calogan}
Michela Paganini, Luke de~Oliveira, and Benjamin Nachman.
\newblock Calogan: Simulating 3d high energy particle showers in multilayer electromagnetic calorimeters with generative adversarial networks.
\newblock \emph{Physical Review D}, 97\penalty0 (1):\penalty0 014021, 2018.

\bibitem[Vallecorsa et~al.(2019)Vallecorsa, Carminati, and Khattak]{vallecorsa20193d}
Sofia Vallecorsa, Federico Carminati, and Gulrukh Khattak.
\newblock 3d convolutional gan for fast simulation.
\newblock In \emph{EPJ Web of Conferences}, volume 214, page 02010. EDP Sciences, 2019.

\bibitem[Erdmann et~al.(2019)Erdmann, Glombitza, and Quast]{erdmann2019precise}
Martin Erdmann, Jonas Glombitza, and Thorben Quast.
\newblock Precise simulation of electromagnetic calorimeter showers using a wasserstein generative adversarial network.
\newblock \emph{Computing and Software for Big Science}, 3:\penalty0 1--13, 2019.

\bibitem[Belayneh et~al.(2020)Belayneh, Carminati, Farbin, Hooberman, Khattak, Liu, Liu, Olivito, Pacela, Pierini, et~al.]{belayneh2020calorimetry}
Dawit Belayneh, Federico Carminati, Amir Farbin, Benjamin Hooberman, Gulrukh Khattak, Miaoyuan Liu, Junze Liu, Dominick Olivito, Vit{\'o}ria~Barin Pacela, Maurizio Pierini, et~al.
\newblock Calorimetry with deep learning: particle simulation and reconstruction for collider physics.
\newblock \emph{The European Physical Journal C}, 80\penalty0 (7):\penalty0 1--31, 2020.

\bibitem[Krause and Shih(2021)]{krause2021caloflow}
Claudius Krause and David Shih.
\newblock Caloflow ii: Even faster and still accurate generation of calorimeter showers with normalizing flows.
\newblock \emph{arXiv preprint arXiv:2110.11377}, 2021.

\bibitem[Bellagente et~al.(2020)Bellagente, Butter, Kasieczka, Plehn, and Winterhalder]{10.21468/SciPostPhys.8.4.070}
Marco Bellagente, Anja Butter, Gregor Kasieczka, Tilman Plehn, and Ramon Winterhalder.
\newblock {How to GAN away detector effects}.
\newblock \emph{SciPost Phys.}, 8:\penalty0 070, 2020.
\newblock \doi{10.21468/SciPostPhys.8.4.070}.
\newblock URL \url{https://scipost.org/10.21468/SciPostPhys.8.4.070}.

\bibitem[De~Oliveira et~al.(2018)De~Oliveira, Paganini, and Nachman]{de2018controlling}
Luke De~Oliveira, Michela Paganini, and Benjamin Nachman.
\newblock Controlling physical attributes in gan-accelerated simulation of electromagnetic calorimeters.
\newblock In \emph{Journal of Physics: Conference Series}, volume 1085, page 042017. IOP Publishing, 2018.

\bibitem[Derkach et~al.(2020)Derkach, Kazeev, Ratnikov, Ustyuzhanin, and Volokhova]{DERKACH2020161804}
Denis Derkach, Nikita Kazeev, Fedor Ratnikov, Andrey Ustyuzhanin, and Alexandra Volokhova.
\newblock Cherenkov detectors fast simulation using neural networks.
\newblock \emph{Nuclear Instruments and Methods in Physics Research Section A: Accelerators, Spectrometers, Detectors and Associated Equipment}, 952:\penalty0 161804, 2020.
\newblock ISSN 0168-9002.
\newblock \doi{https://doi.org/10.1016/j.nima.2019.01.031}.
\newblock URL \url{https://www.sciencedirect.com/science/article/pii/S0168900219300701}.
\newblock 10th International Workshop on Ring Imaging Cherenkov Detectors (RICH 2018).

\bibitem[Alanazi et~al.(2022)Alanazi, Ambrozewicz, Battaglieri, Blin, Kuchera, Li, Liu, McClellan, Melnitchouk, Pritchard, et~al.]{alanazi2022machine}
Y~Alanazi, P~Ambrozewicz, M~Battaglieri, AN~Hiller Blin, MP~Kuchera, Y~Li, T~Liu, RE~McClellan, W~Melnitchouk, E~Pritchard, et~al.
\newblock Machine learning-based event generator for electron-proton scattering.
\newblock \emph{Physical Review D}, 106\penalty0 (9):\penalty0 096002, 2022.

\bibitem[Rehm et~al.(2021)Rehm, Vallecorsa, Saletore, Pabst, Chaibi, Codreanu, Borras, and Kr{\"u}cker]{rehm2021reduced}
Florian Rehm, Sofia Vallecorsa, Vikram Saletore, Hans Pabst, Adel Chaibi, Valeriu Codreanu, Kerstin Borras, and Dirk Kr{\"u}cker.
\newblock Reduced precision strategies for deep learning: a high energy physics generative adversarial network use case.
\newblock \emph{arXiv preprint arXiv:2103.10142}, 2021.

\bibitem[Buhmann et~al.(2022)Buhmann, Diefenbacher, Hundhausen, Kasieczka, Korcari, Eren, Gaede, Kr{\"u}ger, McKeown, and Rustige]{buhmann2022hadrons}
Erik Buhmann, Sascha Diefenbacher, Daniel Hundhausen, Gregor Kasieczka, William Korcari, Engin Eren, Frank Gaede, Katja Kr{\"u}ger, Peter McKeown, and Lennart Rustige.
\newblock Hadrons, better, faster, stronger.
\newblock \emph{Machine Learning: Science and Technology}, 3\penalty0 (2):\penalty0 025014, 2022.

\bibitem[Bieringer et~al.(2022)Bieringer, Butter, Diefenbacher, Eren, Gaede, Hundhausen, Kasieczka, Nachman, Plehn, and Trabs]{bieringer2022calomplification}
Sebastian Bieringer, Anja Butter, Sascha Diefenbacher, Engin Eren, Frank Gaede, Daniel Hundhausen, Gregor Kasieczka, Benjamin Nachman, Tilman Plehn, and Mathias Trabs.
\newblock Calomplification—the power of generative calorimeter models.
\newblock \emph{Journal of Instrumentation}, 17\penalty0 (09):\penalty0 P09028, 2022.

\bibitem[Anderlini et~al.(2023)Anderlini, Barbetti, Derkach, Kazeev, Maevskiy, Mokhnenko, collaboration, et~al.]{anderlini2023towards}
Lucio Anderlini, Matteo Barbetti, Denis Derkach, Nikita Kazeev, Artem Maevskiy, Sergei Mokhnenko, LHCb collaboration, et~al.
\newblock Towards reliable neural generative modeling of detectors.
\newblock In \emph{Journal of Physics: Conference Series}, volume 2438, page 012130. IOP Publishing, 2023.

\bibitem[Hashemi et~al.(2023)Hashemi, Hartmann, Sharifzadeh, Kahn, and Kuhr]{hashemi2023ultra}
Hosein Hashemi, Nikolai Hartmann, Sahand Sharifzadeh, James Kahn, and Thomas Kuhr.
\newblock Ultra-high-resolution detector simulation with intra-event aware gan and self-supervised relational reasoning.
\newblock \emph{arXiv preprint arXiv:2303.08046}, 2023.

\bibitem[Li et~al.(2023)Li, Ostrovskiy, Li, Yang, Al~Kharusi, Anton, Barbeau, Badhrees, Beck, Belov, et~al.]{li2023generative}
S~Li, I~Ostrovskiy, Z~Li, L~Yang, S~Al~Kharusi, G~Anton, PS~Barbeau, I~Badhrees, D~Beck, V~Belov, et~al.
\newblock Generative adversarial networks for scintillation signal simulation in exo-200.
\newblock \emph{Journal of Instrumentation}, 18\penalty0 (06):\penalty0 P06005, 2023.

\bibitem[Lu et~al.(2021)Lu, Collado, Whiteson, and Baldi]{lu2021sparse}
Yadong Lu, Julian Collado, Daniel Whiteson, and Pierre Baldi.
\newblock Sparse autoregressive models for scalable generation of sparse images in particle physics.
\newblock \emph{Physical Review D}, 103\penalty0 (3):\penalty0 036012, 2021.

\bibitem[Butter et~al.(2023{\natexlab{a}})Butter, Heimel, Hummerich, Krebs, Plehn, Rousselot, and Vent]{butter2023generative}
Anja Butter, Theo Heimel, Sander Hummerich, Tobias Krebs, Tilman Plehn, Armand Rousselot, and Sophia Vent.
\newblock Generative networks for precision enthusiasts.
\newblock \emph{SciPost Physics}, 14\penalty0 (4):\penalty0 078, 2023{\natexlab{a}}.

\bibitem[Diefenbacher et~al.(2023{\natexlab{a}})Diefenbacher, Eren, Gaede, Kasieczka, Krause, Shekhzadeh, and Shih]{diefenbacher2023l2lflows}
Sascha Diefenbacher, Engin Eren, Frank Gaede, Gregor Kasieczka, Claudius Krause, Imahn Shekhzadeh, and David Shih.
\newblock L2lflows: Generating high-fidelity 3d calorimeter images.
\newblock \emph{arXiv preprint arXiv:2302.11594}, 2023{\natexlab{a}}.

\bibitem[Raine et~al.(2023)Raine, Leigh, Zoch, and Golling]{raine2023nu}
John~Andrew Raine, Matthew Leigh, Knut Zoch, and Tobias Golling.
\newblock {$nu^2$}-flows: Fast and improved neutrino reconstruction in multi-neutrino final states with conditional normalizing flows.
\newblock \emph{arXiv preprint arXiv:2307.02405}, 2023.

\bibitem[Golling et~al.(2023)Golling, Kasieczka, Krause, Mastandrea, Nachman, Raine, Sengupta, Shih, and Sommerhalder]{golling2023interplay}
Tobias Golling, Gregor Kasieczka, Claudius Krause, Radha Mastandrea, Benjamin Nachman, John~Andrew Raine, Debajyoti Sengupta, David Shih, and Manuel Sommerhalder.
\newblock The interplay of machine learning--based resonant anomaly detection methods.
\newblock \emph{arXiv preprint arXiv:2307.11157}, 2023.

\bibitem[Xu et~al.(2023)Xu, Han, Ju, and Wang]{xu2023generative}
Allison Xu, Shuo Han, Xiangyang Ju, and Haichen Wang.
\newblock Generative machine learning for detector response modeling with a conditional normalizing flow.
\newblock \emph{arXiv preprint arXiv:2303.10148}, 2023.

\bibitem[Krause et~al.(2023)Krause, Nachman, Pang, Shih, and Zhu]{krause2023anomaly}
Claudius Krause, Benjamin Nachman, Ian Pang, David Shih, and Yunhao Zhu.
\newblock Anomaly detection with flow-based fast calorimeter simulators.
\newblock \emph{arXiv preprint arXiv:2312.11618}, 2023.

\bibitem[Lutkus et~al.(2022)Lutkus, Wongjirad, and Aeron]{lutkus2022towards}
Paul Lutkus, Taritree Wongjirad, and Shuchin Aeron.
\newblock Towards designing and exploiting generative networks for neutrino physics experiments using liquid argon time projection chambers.
\newblock \emph{arXiv preprint arXiv:2204.02496}, 2022.

\bibitem[Rombach et~al.(2022)Rombach, Blattmann, Lorenz, Esser, and Ommer]{rombach2022high}
Robin Rombach, Andreas Blattmann, Dominik Lorenz, Patrick Esser, and Bj{\"{o}}rn Ommer.
\newblock High-resolution image synthesis with latent diffusion models.
\newblock In \emph{{CVPR}}, pages 10674--10685. {IEEE}, 2022.

\bibitem[Yi et~al.(2021)Yi, Guo, and Bai]{yi2021exploring}
Da~Yi, Chao Guo, and Tianxiang Bai.
\newblock Exploring painting synthesis with diffusion models.
\newblock In \emph{2021 IEEE 1st International Conference on Digital Twins and Parallel Intelligence (DTPI)}, pages 332--335. IEEE, 2021.

\bibitem[Peng et~al.(2023)Peng, Zhao, Xie, Fukusato, and Miyata]{peng2023difffacesketch}
Yichen Peng, Chunqi Zhao, Haoran Xie, Tsukasa Fukusato, and Kazunori Miyata.
\newblock Difffacesketch: High-fidelity face image synthesis with sketch-guided latent diffusion model.
\newblock \emph{arXiv preprint arXiv:2302.06908}, 2023.

\bibitem[Leigh et~al.(2023)Leigh, Sengupta, Qu{\'e}tant, Raine, Zoch, and Golling]{leigh2023pc}
Matthew Leigh, Debajyoti Sengupta, Guillaume Qu{\'e}tant, John~Andrew Raine, Knut Zoch, and Tobias Golling.
\newblock Pc-jedi: Diffusion for particle cloud generation in high energy physics.
\newblock \emph{arXiv preprint arXiv:2303.05376}, 2023.

\bibitem[Diefenbacher et~al.(2023{\natexlab{b}})Diefenbacher, Liu, Mikuni, Nachman, and Nie]{diefenbacher2023improving}
Sascha Diefenbacher, Guan-Horng Liu, Vinicius Mikuni, Benjamin Nachman, and Weili Nie.
\newblock Improving generative model-based unfolding with schrödinger bridges.
\newblock \emph{arXiv preprint arXiv:2308.12351}, 2023{\natexlab{b}}.

\bibitem[Butter et~al.(2023{\natexlab{b}})Butter, Huetsch, Schweitzer, Plehn, Sorrenson, and Spinner]{butter2023jet}
Anja Butter, Nathan Huetsch, Sofia~Palacios Schweitzer, Tilman Plehn, Peter Sorrenson, and Jonas Spinner.
\newblock Jet diffusion versus jetgpt--modern networks for the lhc.
\newblock \emph{arXiv preprint arXiv:2305.10475}, 2023{\natexlab{b}}.

\bibitem[Mikuni and Nachman(2022)]{vinicius2022Calorimeter}
Vinicius Mikuni and Benjamin Nachman.
\newblock Score-based generative models for calorimeter shower simulation.
\newblock \emph{Phys. Rev. D}, 106:\penalty0 092009, Nov 2022.
\newblock \doi{10.1103/PhysRevD.106.092009}.
\newblock URL \url{https://link.aps.org/doi/10.1103/PhysRevD.106.092009}.

\bibitem[Buhmann et~al.(2023)Buhmann, Gaede, Kasieczka, Korol, Korcari, Kr{\"u}ger, and McKeown]{buhmann2023caloclouds}
Erik Buhmann, Frank Gaede, Gregor Kasieczka, Anatolii Korol, William Korcari, Katja Kr{\"u}ger, and Peter McKeown.
\newblock Caloclouds ii: Ultra-fast geometry-independent highly-granular calorimeter simulation.
\newblock \emph{arXiv preprint arXiv:2309.05704}, 2023.

\bibitem[Acosta et~al.(2023)Acosta, Mikuni, Nachman, Arratia, Barish, Karki, Milton, Karande, and Angerami]{acosta2023comparison}
Fernando~Torales Acosta, Vinicius Mikuni, Benjamin Nachman, Miguel Arratia, Kenneth Barish, Bishnu Karki, Ryan Milton, Piyush Karande, and Aaron Angerami.
\newblock Comparison of point cloud and image-based models for calorimeter fast simulation.
\newblock \emph{arXiv preprint arXiv:2307.04780}, 2023.

\bibitem[Song et~al.(2021)Song, Sohl-Dickstein, Kingma, Kumar, Ermon, and Poole]{song2021scorebased}
Yang Song, Jascha Sohl-Dickstein, Diederik~P Kingma, Abhishek Kumar, Stefano Ermon, and Ben Poole.
\newblock Score-based generative modeling through stochastic differential equations.
\newblock In \emph{International Conference on Learning Representations}, 2021.
\newblock URL \url{https://openreview.net/forum?id=PxTIG12RRHS}.

\bibitem[Bond-Taylor et~al.(2021)Bond-Taylor, Leach, Long, and Willcocks]{bond2021deep}
Sam Bond-Taylor, Adam Leach, Yang Long, and Chris~G Willcocks.
\newblock Deep generative modelling: A comparative review of vaes, gans, normalizing flows, energy-based and autoregressive models.
\newblock \emph{IEEE transactions on pattern analysis and machine intelligence}, 2021.

\bibitem[Yang et~al.(2024)Yang, Zhang, Song, Hong, Xu, Zhao, Zhang, Cui, and Yang]{yang2023diffusion}
Ling Yang, Zhilong Zhang, Yang Song, Shenda Hong, Runsheng Xu, Yue Zhao, Wentao Zhang, Bin Cui, and Ming{-}Hsuan Yang.
\newblock Diffusion models: {A} comprehensive survey of methods and applications.
\newblock \emph{{ACM} Comput. Surv.}, 56\penalty0 (4):\penalty0 105:1--105:39, 2024.

\bibitem[Chen et~al.(2023{\natexlab{a}})Chen, Lee, and Lu]{chen2023improved}
Hongrui Chen, Holden Lee, and Jianfeng Lu.
\newblock Improved analysis of score-based generative modeling: User-friendly bounds under minimal smoothness assumptions.
\newblock In \emph{International Conference on Machine Learning}, pages 4735--4763, Honolulu, Hawaii, USA, 2023{\natexlab{a}}. PMLR.

\bibitem[Chen et~al.(2023{\natexlab{b}})Chen, Chewi, Li, Li, Salim, and Zhang]{chen2022sampling}
Sitan Chen, Sinho Chewi, Jerry Li, Yuanzhi Li, Adil Salim, and Anru Zhang.
\newblock Sampling is as easy as learning the score: theory for diffusion models with minimal data assumptions.
\newblock In \emph{The Eleventh International Conference on Learning Representations}, Kigali, Rwanda, 2023{\natexlab{b}}.
\newblock URL \url{https://openreview.net/forum?id=zyLVMgsZ0U_}.

\bibitem[De~Bortoli(2022)]{de2022convergence}
Valentin De~Bortoli.
\newblock Convergence of denoising diffusion models under the manifold hypothesis.
\newblock \emph{Transactions on Machine Learning Research}, 2022.
\newblock ISSN 2835-8856.
\newblock URL \url{https://openreview.net/forum?id=MhK5aXo3gB}.
\newblock Expert Certification.

\bibitem[Zhang and Chen(2023)]{zhang2022fast}
Qinsheng Zhang and Yongxin Chen.
\newblock Fast sampling of diffusion models with exponential integrator.
\newblock In \emph{The Eleventh International Conference on Learning Representations}, Kigali, Rwanda, 2023.
\newblock URL \url{https://openreview.net/forum?id=Loek7hfb46P}.

\bibitem[Anderson(1982)]{anderson1982reverse}
Brian~DO Anderson.
\newblock Reverse-time diffusion equation models.
\newblock \emph{Stochastic Processes and their Applications}, 12\penalty0 (3):\penalty0 313--326, 1982.

\bibitem[Conforti and Léonard(2022)]{Conforti_2022}
Giovanni Conforti and Christian Léonard.
\newblock Time reversal of markov processes with jumps under a finite entropy condition.
\newblock \emph{Stochastic Processes and their Applications}, 144:\penalty0 85--124, 2022.
\newblock ISSN 0304-4149.
\newblock \doi{https://doi.org/10.1016/j.spa.2021.10.002}.
\newblock URL \url{https://www.sciencedirect.com/science/article/pii/S0304414921001654}.

\bibitem[Hyv{{\"a}}rinen(2005)]{hyvarinen2005estimation}
Aapo Hyv{{\"a}}rinen.
\newblock Estimation of non-normalized statistical models by score matching.
\newblock \emph{Journal of Machine Learning Research}, 6\penalty0 (24):\penalty0 695--709, 2005.
\newblock URL \url{http://jmlr.org/papers/v6/hyvarinen05a.html}.

\bibitem[Adams et~al.(2020)Adams, Terao, and Wongjirad]{adams2020pilarnet}
Corey Adams, Kazuhiro Terao, and Taritree Wongjirad.
\newblock Pilarnet: public dataset for particle imaging liquid argon detectors in high energy physics.
\newblock \emph{arXiv preprint arXiv:2006.01993}, 2020.

\bibitem[Krizhevsky(2009)]{krizhevsky2009learning}
A.~Krizhevsky.
\newblock {Learning Multiple Layers of Features from Tiny Images}.
\newblock Technical report, Univ. Toronto, Toronto, ON, Canada, 2009.

\bibitem[Collaboration et~al.(2019)Collaboration, Adams, Alrashed, An, Anthony, Asaadi, Ashkenazi, Auger, Balasubramanian, Baller, et~al.]{collaboration2019deep}
MicroBooNE Collaboration, C~Adams, M~Alrashed, R~An, J~Anthony, J~Asaadi, A~Ashkenazi, M~Auger, S~Balasubramanian, B~Baller, et~al.
\newblock Deep neural network for pixel-level electromagnetic particle identification in the microboone liquid argon time projection chamber.
\newblock \emph{Physical Review D}, 99\penalty0 (9):\penalty0 092001, 2019.

\bibitem[Abratenko et~al.(2021)Abratenko, Alrashed, An, Anthony, Asaadi, Ashkenazi, Balasubramanian, Baller, Barnes, Barr, et~al.]{abratenko2021semantic}
P~Abratenko, M~Alrashed, R~An, J~Anthony, J~Asaadi, A~Ashkenazi, S~Balasubramanian, B~Baller, C~Barnes, G~Barr, et~al.
\newblock Semantic segmentation with a sparse convolutional neural network for event reconstruction in microboone.
\newblock \emph{Physical Review D}, 103\penalty0 (5):\penalty0 052012, 2021.

\bibitem[Abratenko et~al.(2022)Abratenko, An, Anthony, Arellano, Asaadi, Ashkenazi, Balasubramanian, Baller, Barnes, Barr, et~al.]{abratenko2022search}
P~Abratenko, R~An, J~Anthony, L~Arellano, J~Asaadi, A~Ashkenazi, S~Balasubramanian, B~Baller, C~Barnes, G~Barr, et~al.
\newblock Search for an anomalous excess of charged-current quasielastic $\nu$ e interactions with the microboone experiment using deep-learning-based reconstruction.
\newblock \emph{Physical Review D}, 105\penalty0 (11):\penalty0 112003, 2022.

\bibitem[Gretton et~al.(2012)Gretton, Borgwardt, Rasch, Sch{{\"o}}lkopf, and Smola]{JMLR:v13:gretton12a}
Arthur Gretton, Karsten~M. Borgwardt, Malte~J. Rasch, Bernhard Sch{{\"o}}lkopf, and Alexander Smola.
\newblock A kernel two-sample test.
\newblock \emph{Journal of Machine Learning Research}, 13\penalty0 (25):\penalty0 723--773, 2012.
\newblock ISSN 1532-4435.
\newblock URL \url{http://jmlr.org/papers/v13/gretton12a.html}.

\bibitem[Genevay et~al.(2018)Genevay, Peyr{\'e}, and Cuturi]{genevay2018learning}
Aude Genevay, Gabriel Peyr{\'e}, and Marco Cuturi.
\newblock Learning generative models with sinkhorn divergences.
\newblock In Amos Storkey and Fernando Perez-Cruz, editors, \emph{Proceedings of the Twenty-First International Conference on Artificial Intelligence and Statistics}, volume~84 of \emph{Proceedings of Machine Learning Research}, pages 1608--1617, Stockholm, Sweden, 09--11 Apr 2018. PMLR.
\newblock URL \url{https://proceedings.mlr.press/v84/genevay18a.html}.

\bibitem[Mena and Niles{-}Weed(2019)]{mena2019statistical}
Gonzalo Mena and Jonathan Niles{-}Weed.
\newblock Statistical bounds for entropic optimal transport: sample complexity and the central limit theorem.
\newblock In Hanna~M. Wallach, Hugo Larochelle, Alina Beygelzimer, Florence d'Alch{\'{e}}{-}Buc, Emily~B. Fox, and Roman Garnett, editors, \emph{Advances in Neural Information Processing Systems 32: Annual Conference on Neural Information Processing Systems 2019, NeurIPS 2019, December 8-14, 2019, Vancouver, BC, Canada}, pages 4543--4553, Vancouver, BC, Canada, 2019.
\newblock URL \url{https://proceedings.neurips.cc/paper/2019/hash/5acdc9ca5d99ae66afdfe1eea0e3b26b-Abstract.html}.

\bibitem[Sriperumbudur et~al.(2012)Sriperumbudur, Fukumizu, Gretton, Sch{\"o}lkopf, and Lanckriet]{Sriperumbudur_2012}
Bharath~K. Sriperumbudur, Kenji Fukumizu, Arthur Gretton, Bernhard Sch{\"o}lkopf, and Gert R.~G. Lanckriet.
\newblock {On the empirical estimation of integral probability metrics}.
\newblock \emph{Electronic Journal of Statistics}, 6\penalty0 (none):\penalty0 1550 -- 1599, 2012.
\newblock \doi{10.1214/12-EJS722}.
\newblock URL \url{https://doi.org/10.1214/12-EJS722}.

\bibitem[Genevay et~al.(2019)Genevay, Chizat, Bach, Cuturi, and Peyr\'{e}]{pmlr-v89-genevay19a}
Aude Genevay, L\'{e}na\"{i}c Chizat, Francis Bach, Marco Cuturi, and Gabriel Peyr\'{e}.
\newblock Sample complexity of sinkhorn divergences.
\newblock In Kamalika Chaudhuri and Masashi Sugiyama, editors, \emph{Proceedings of the Twenty-Second International Conference on Artificial Intelligence and Statistics}, volume~89 of \emph{Proceedings of Machine Learning Research}, pages 1574--1583. PMLR, 16--18 Apr 2019.
\newblock URL \url{https://proceedings.mlr.press/v89/genevay19a.html}.

\bibitem[Heusel et~al.(2017)Heusel, Ramsauer, Unterthiner, Nessler, and Hochreiter]{FIDpaper}
Martin Heusel, Hubert Ramsauer, Thomas Unterthiner, Bernhard Nessler, and Sepp Hochreiter.
\newblock Gans trained by a two time-scale update rule converge to a local nash equilibrium.
\newblock In \emph{Proceedings of the 31st International Conference on Neural Information Processing Systems}, volume~30 of \emph{NIPS'17}, page 6629–6640, Red Hook, NY, USA, 2017. Curran Associates Inc.
\newblock ISBN 9781510860964.

\bibitem[Szegedy et~al.(2016)Szegedy, Vanhoucke, Ioffe, Shlens, and Wojna]{szegedy2016rethinking}
C.~Szegedy, V.~Vanhoucke, S.~Ioffe, J.~Shlens, and Z.~Wojna.
\newblock Rethinking the inception architecture for computer vision.
\newblock In \emph{2016 IEEE Conference on Computer Vision and Pattern Recognition (CVPR)}, pages 2818--2826, Los Alamitos, CA, USA, jun 2016. IEEE Computer Society.
\newblock \doi{10.1109/CVPR.2016.308}.
\newblock URL \url{https://doi.ieeecomputersociety.org/10.1109/CVPR.2016.308}.

\bibitem[Ester et~al.(1996)Ester, Kriegel, Sander, and Xu]{ester1996density}
Martin Ester, Hans-Peter Kriegel, J\"{o}rg Sander, and Xiaowei Xu.
\newblock A density-based algorithm for discovering clusters in large spatial databases with noise.
\newblock In \emph{Proceedings of the Second International Conference on Knowledge Discovery and Data Mining}, KDD'96, page 226–231. AAAI Press, 1996.

\bibitem[Komiske et~al.(2019)Komiske, Metodiev, and Thaler]{thaler2019physicsemd}
Patrick~T. Komiske, Eric~M. Metodiev, and Jesse Thaler.
\newblock Metric space of collider events.
\newblock \emph{Phys. Rev. Lett.}, 123:\penalty0 041801, Jul 2019.
\newblock \doi{10.1103/PhysRevLett.123.041801}.
\newblock URL \url{https://link.aps.org/doi/10.1103/PhysRevLett.123.041801}.

\bibitem[Zheng et~al.(2023)Zheng, Nie, Vahdat, Azizzadenesheli, and Anandkumar]{zheng2023fast}
Hongkai Zheng, Weili Nie, Arash Vahdat, Kamyar Azizzadenesheli, and Anima Anandkumar.
\newblock Fast sampling of diffusion models via operator learning.
\newblock In Andreas Krause, Emma Brunskill, Kyunghyun Cho, Barbara Engelhardt, Sivan Sabato, and Jonathan Scarlett, editors, \emph{Proceedings of the 40th International Conference on Machine Learning}, volume 202 of \emph{Proceedings of Machine Learning Research}, pages 42390--42402, Honolulu, Hawaii, USA, 23--29 Jul 2023. PMLR.
\newblock URL \url{https://proceedings.mlr.press/v202/zheng23d.html}.

\bibitem[Khrulkov et~al.(2022)Khrulkov, Ryzhakov, Chertkov, and Oseledets]{khrulkov2022understanding}
Valentin Khrulkov, Gleb Ryzhakov, Andrei Chertkov, and Ivan Oseledets.
\newblock Understanding ddpm latent codes through optimal transport.
\newblock \emph{arXiv preprint arXiv:2202.07477}, 2022.

\bibitem[De~Bortoli et~al.(2021)De~Bortoli, Thornton, Heng, and Doucet]{de2021diffusion}
Valentin De~Bortoli, James Thornton, Jeremy Heng, and Arnaud Doucet.
\newblock Diffusion schr\"{o}dinger bridge with applications to score-based generative modeling.
\newblock In M.~Ranzato, A.~Beygelzimer, Y.~Dauphin, P.S. Liang, and J.~Wortman Vaughan, editors, \emph{Advances in Neural Information Processing Systems}, volume~34, pages 17695--17709. Curran Associates, Inc., 2021.
\newblock URL \url{https://proceedings.neurips.cc/paper_files/paper/2021/file/940392f5f32a7ade1cc201767cf83e31-Paper.pdf}.

\bibitem[Liu et~al.(2023)Liu, Luo, Xu, Jaakkola, and Tegmark]{liu2023genphys}
Ziming Liu, Di~Luo, Yilun Xu, Tommi Jaakkola, and Max Tegmark.
\newblock Genphys: From physical processes to generative models.
\newblock \emph{arXiv preprint arXiv:2304.02637}, 2023.

\end{thebibliography}
\bibliographystyle{unsrtnat}

\clearpage

\appendix

\section{Additional Details for Training and Generating Images}
\label{appendix:traingen_details}

The realization of a generative diffusion model involves many possible choices.
In order to facilitate reproduction and exploration in future work, we document all of the parameters used in the work presented in Table \ref{tab:configs}
These parameters are specific to the code repository by Song et al.~\cite{song2021scorebased}, which we used for training and generation.
We left the majority of parameters unchanged from the default CIFAR-10 configurations. 
We leave the exploration of hyperparameters turning to later work.

\section{Additional Details for Our Evaluation Metrics}
\label{appendix:eval_details}

Our high-dimensional goodness of fit tests required specifying parameters. 
For MMD, we used a mixed kernel with sigma values: $2^{10}, 2^{11}, 2^{12}, 2^{13}, 2^{14}, \text{and } 2^{15}$.
We found that smaller sigma values had less ability to differentiate between epochs. 
Larger sigma values vertically scaled the distances without affecting the shape of the distribution. 
For Sinkhorn Divergence, we used eps = 1, alternate eps values primarily affected vertical scaling.

Table~\ref{appendix:SSNet-FID_values} provides values for future comparisons of the SSNet-FID score images generated after training for a different number of epochs.

\vfill\break 

\section{Additional Images for Similarity Measures Used for Studying Potential Mode Collapse}
\label{appendix:moreimages_mode_check}

\begin{figure*}[h!]
    \centering
    \includegraphics[page=1,width=0.85\textwidth]{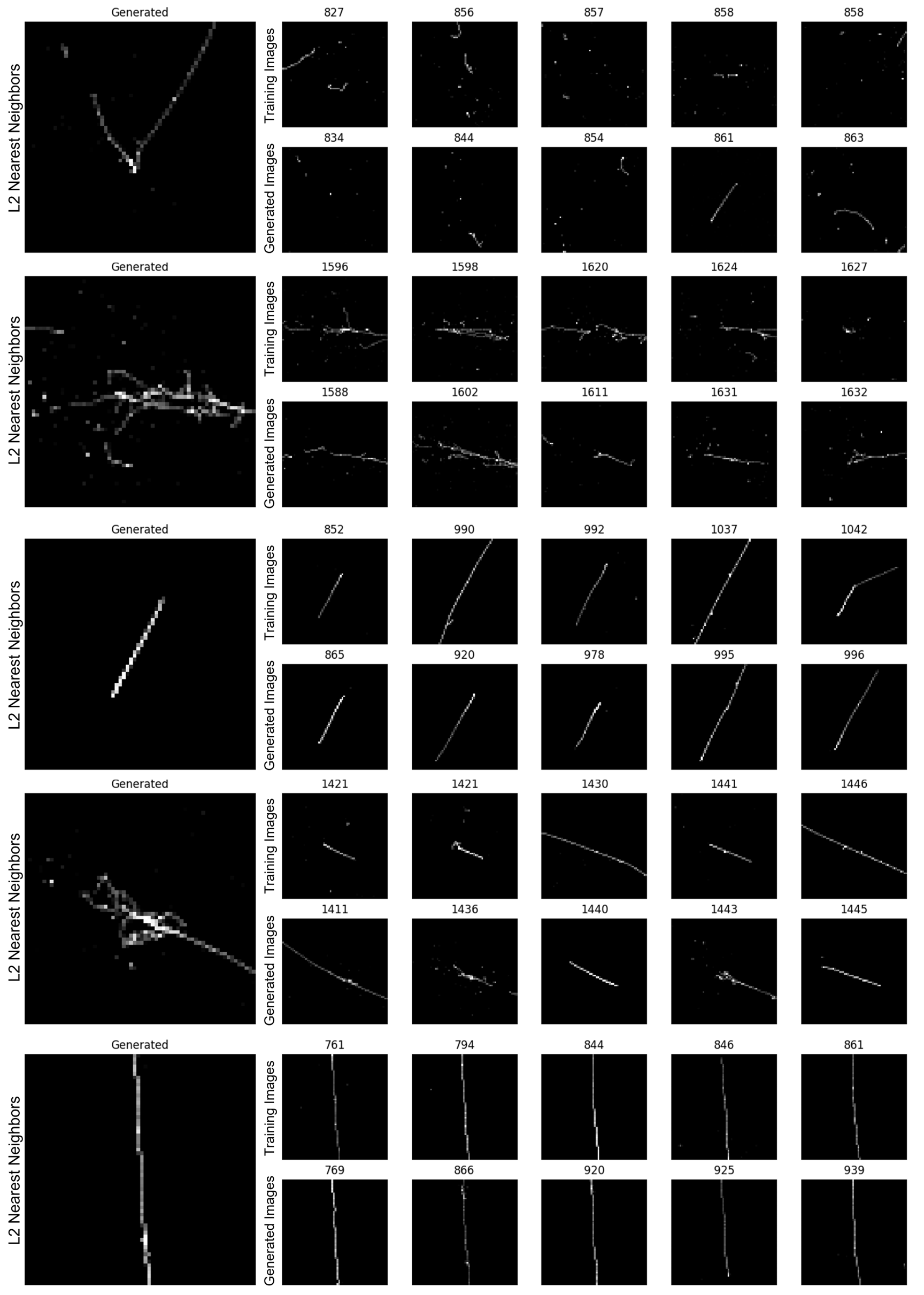}
    \caption{Checking for mode collapse by finding the nearest neighbors using Euclidian distance (L2 norm). The five large images are generated events from 50 epochs of training. For each generated image we have found the five closest matches from the training dataset (top row) and generated dataset (bottom row). We have excluded itself (distance zero) in the generated image search. The number above each comparison image is its distance from the comparison image. There are an equal number of images (50,000) in the training and generated datasets. We see that images are distinct and have reasonably large distances, suggesting that we are generating unique images.} 
    \label{fig:l2_neighbors}
\end{figure*}

\begin{figure*}[h!]
    \centering
    \includegraphics[page=1,width=0.85\textwidth]{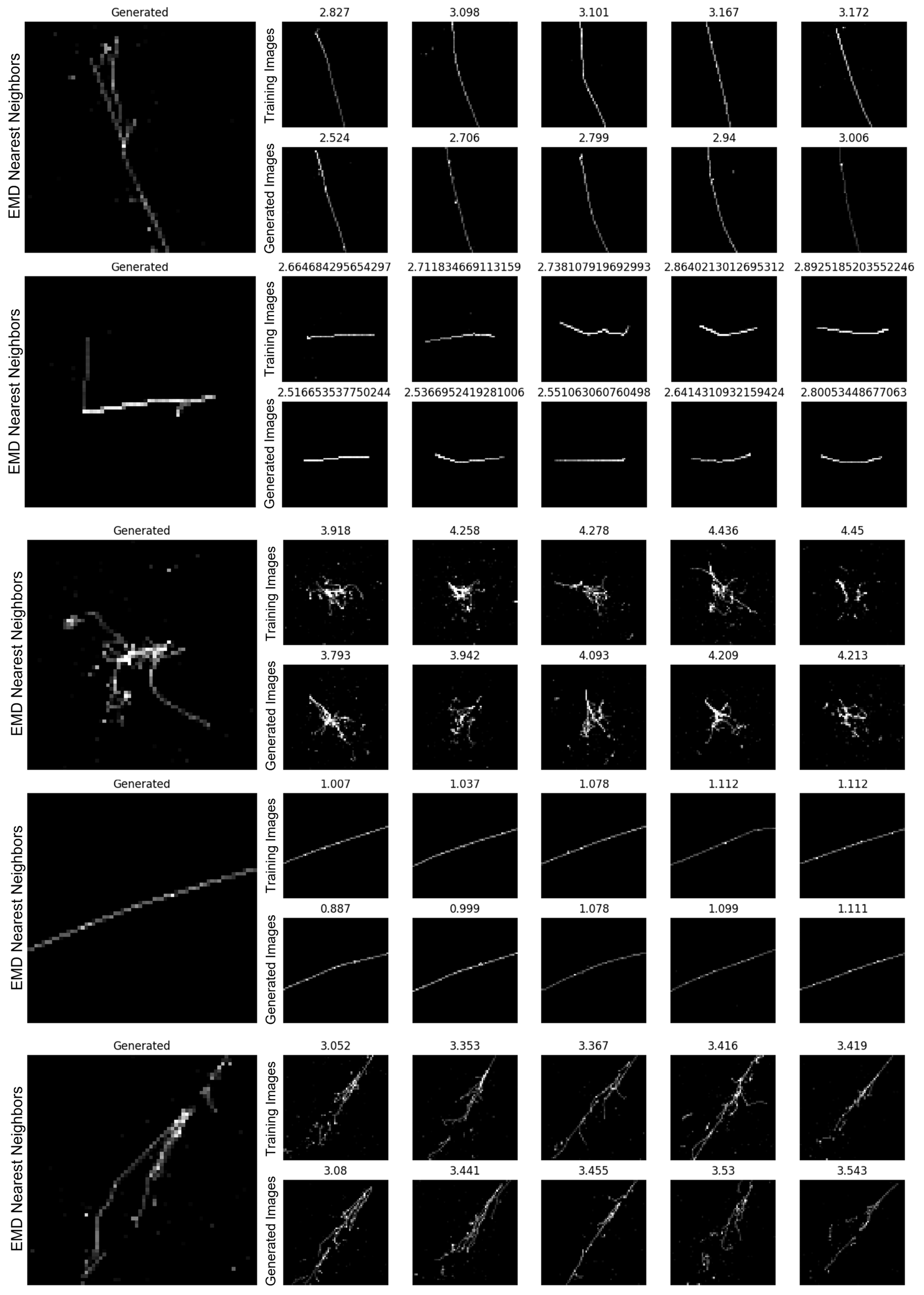}
    \caption{Checking for mode collapse by finding the nearest neighbors using Earth (Energy) Mover's Distance. The five large images are generated events from 50 epochs of training. For each generated image we have found the five closest matches from the training dataset (top row) and generated dataset (bottom row). We have excluded itself (distance zero) in the generated image search. The number above each comparison image is its distance from the comparison image. There are an equal number of images (50,000) in the training and generated datasets. We see that images are distinct and have reasonably large distances, suggesting that we are generating unique images.}
    \label{fig:emd_neighbors}
\end{figure*}

Figures \ref{fig:l2_neighbors} and \ref{fig:emd_neighbors} each provide five examples of nearest neighbors for a generated image in the training and generated image datasets. The distances are calculated using l2 Euclidean norm and Earth Mover's Distance (EMD), respectively. 


\section{Additional Uncurated Examples of Training and Generated Images}
\label{appendix:moretrain_gen_images}

\begin{figure*}[h]
    \centering
    \includegraphics[page=1,width=0.9\textwidth]{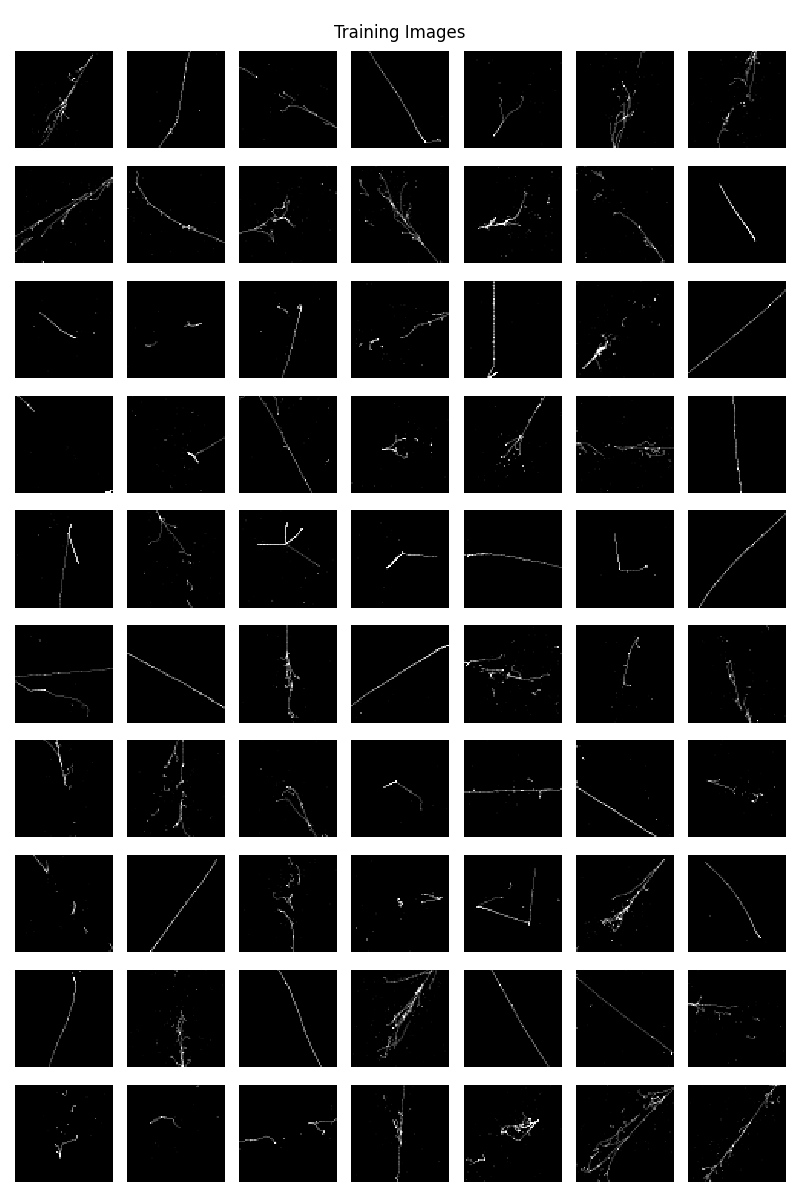}
    \caption{Randomly selected images from the training dataset.}
    \label{fig:training_image_page}
\end{figure*}

\begin{figure*}[h]
    \centering
    \includegraphics[page=1,width=0.9\textwidth]{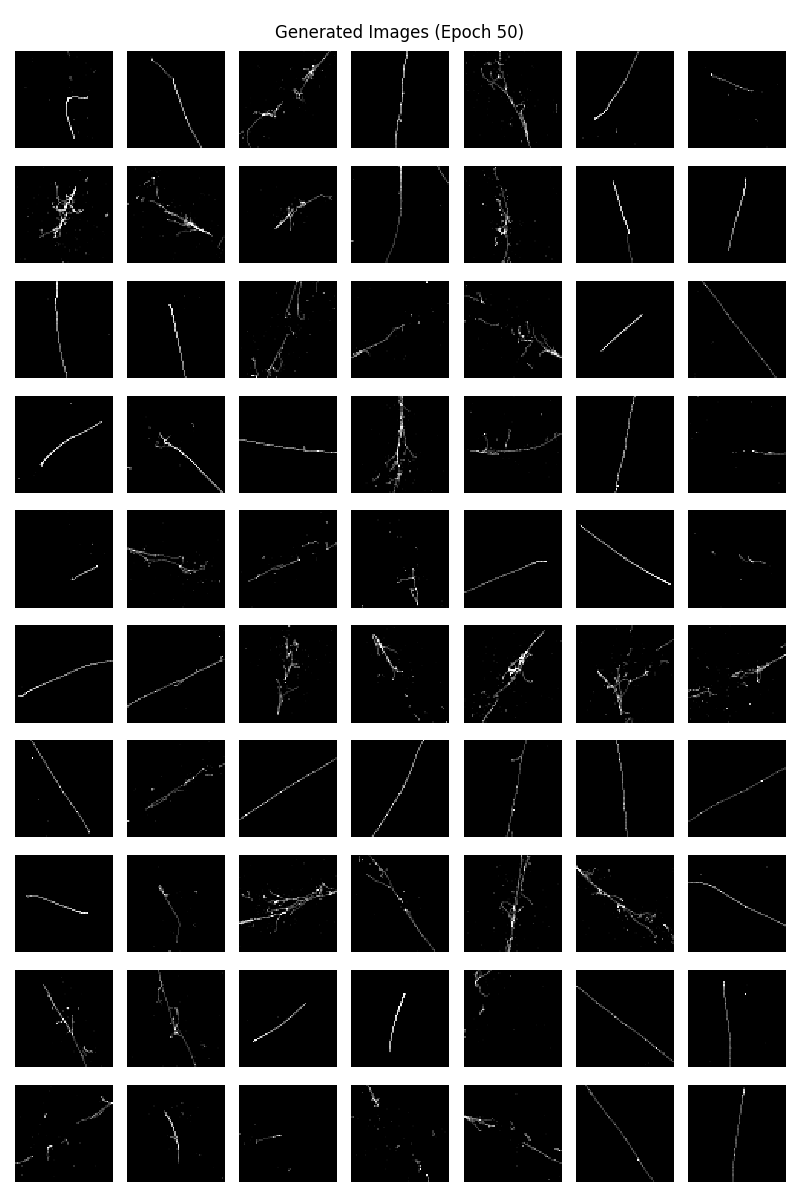}
    \caption{Randomly selected images from the epoch 50 generated image dataset.}
    \label{fig:generated_image_page}
\end{figure*}

Figures~\ref{fig:training_image_page} and Figures~\ref{fig:generated_image_page} include more examples of training and generated images, respectively.

\section{VQ-VAE Comparison}
\label{appendix:VQ-VAE}

\begin{figure*}[h]
    \centering
    \includegraphics[width=0.85\textwidth]{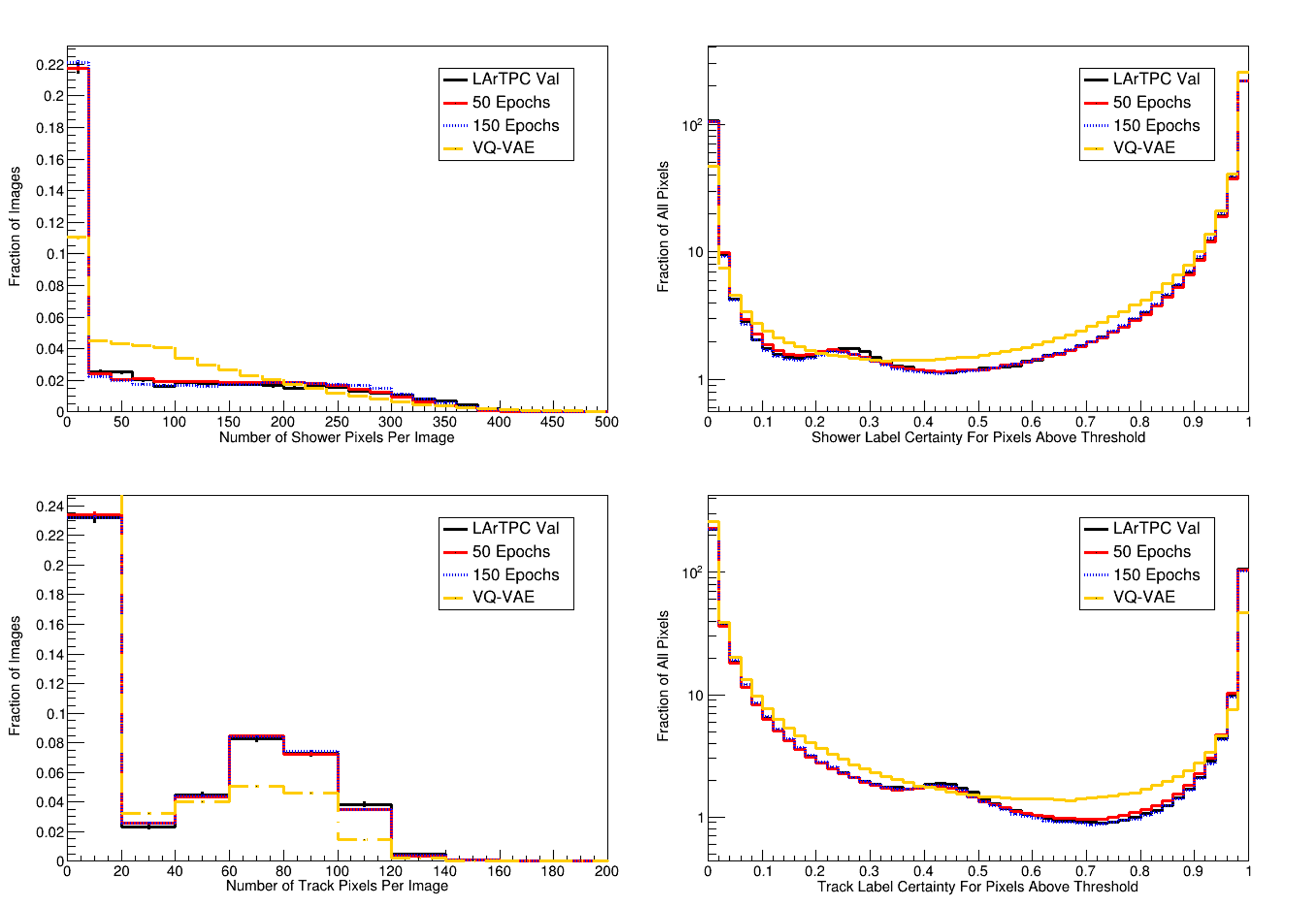} 
    \caption{SSNet Pixel Value Histograms with VQVAE.}
    \label{fig:SSNet_hists_vqvae}
\end{figure*}

\begin{figure*}
\centering
\begin{subfigure}[b]{0.4\textwidth}
    \centering
    \includegraphics[width=\textwidth]{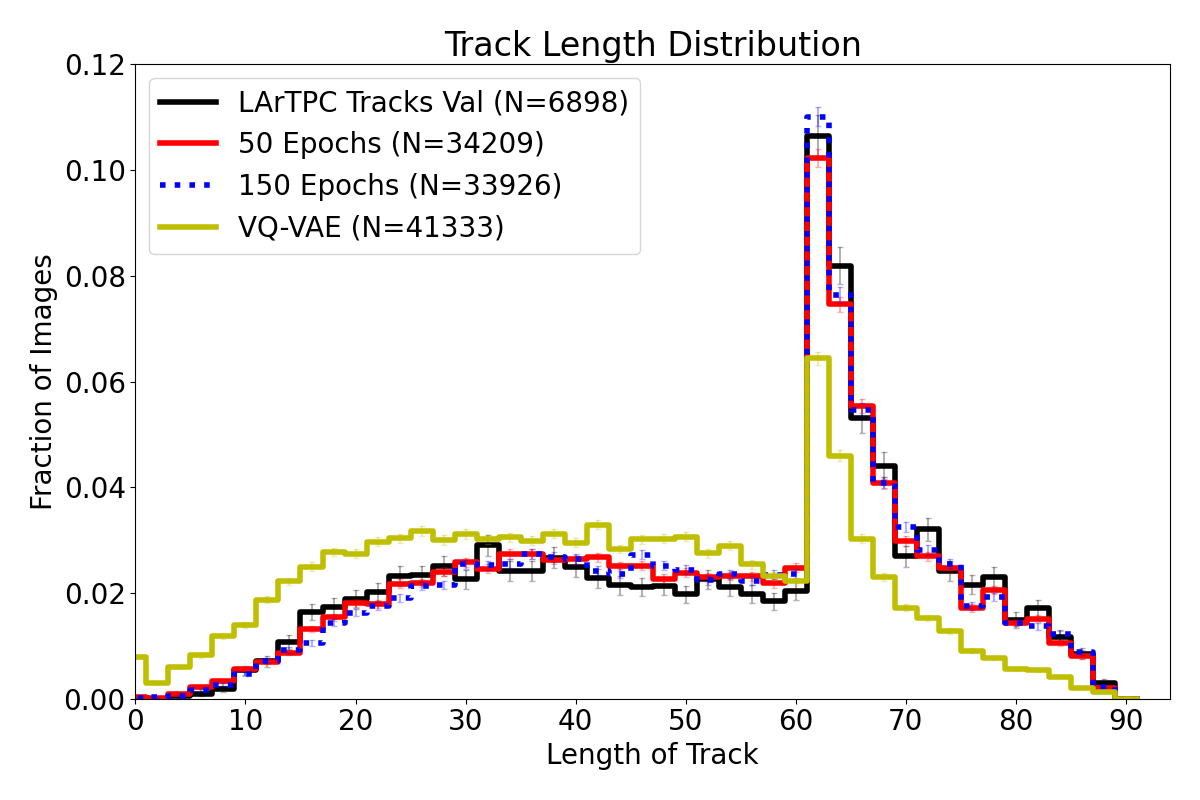}
    \caption{Track length with VQ-VAE.}
    \label{fig:track_length_vqvae}
\end{subfigure}
\hfill
\begin{subfigure}[b]{0.4\textwidth}
    \centering
    \includegraphics[width=\textwidth]{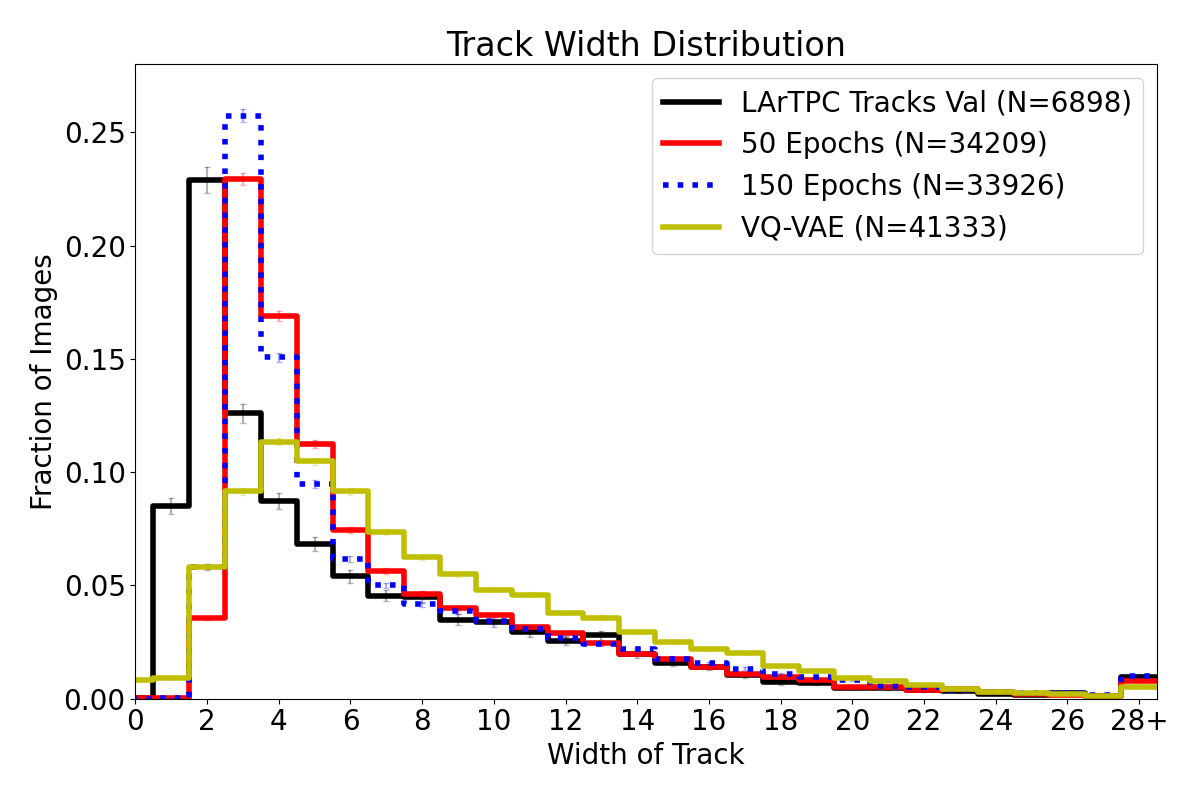}
    \caption{Track width with VQ-VAE.}
    \label{fig:track_width_vqvae}
\end{subfigure}
\begin{subfigure}[b]{0.4\textwidth}
    \centering
    \includegraphics[width=\textwidth]{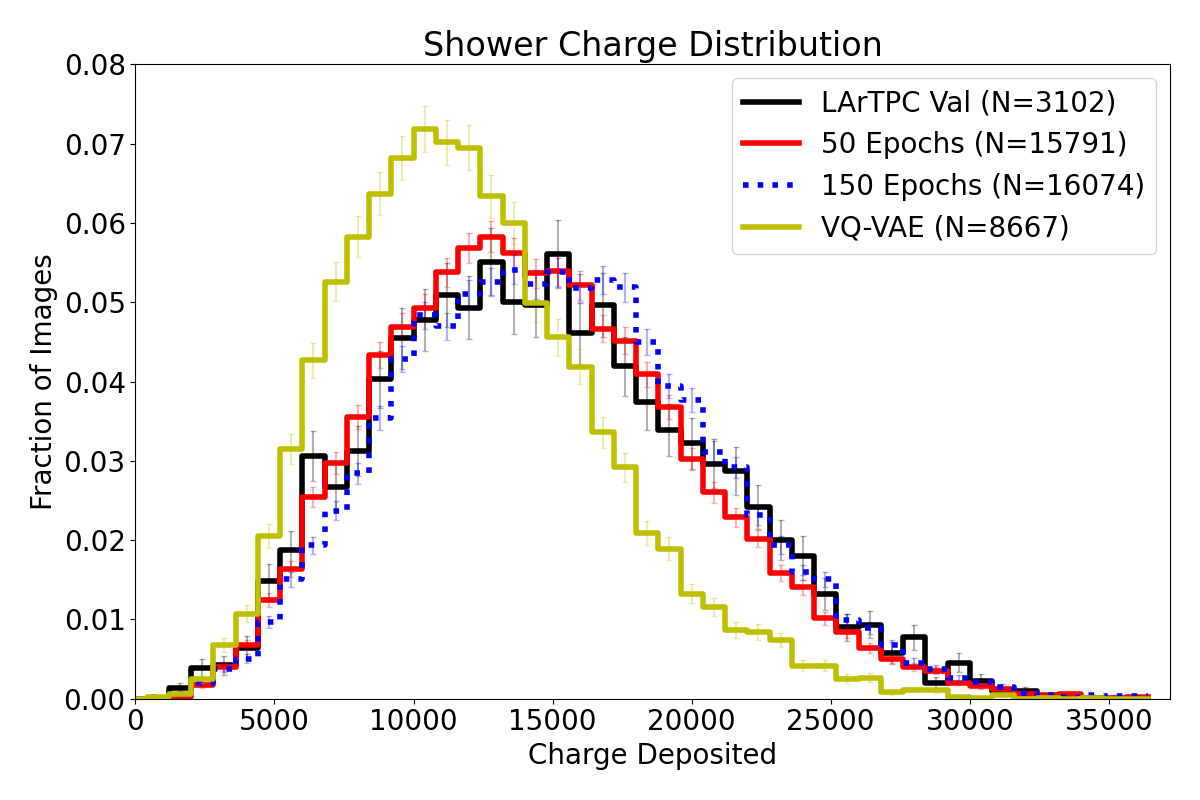}
    \caption{Shower charge with VQ-VAE.}
    \label{fig:shower_charge_vqvae}
\end{subfigure}
\caption{Physics-based metrics with VQ-VAE.}
\label{fig:physics_vqvae}
\end{figure*}

We revisited our previous attempt \cite{lutkus2022towards} using a vector quantized variational autoencoder (VQ-VAE) model to generate LArTPC images. 
We generated 50,000 new images to compare performance against our diffusion model and ran the images through SSNet and our physics-based metrics.  
These results can be found in Figures~\ref{fig:SSNet_hists_vqvae} and~\ref{fig:physics_vqvae}, respectively.
While VQ-VAE performs better than expected from our previous analysis, it is still noticeably worse than epochs 50 and 150 across all our metrics. 




\clearpage

\begin{table*}[h]
\begin{minipage}{0.4\textwidth}
\begin{tabular}{|l|l|}
\hline
\multicolumn{2}{|c|}{\bf{default\_particle\_configs.py}} \\ \hline
\multicolumn{2}{|c|}{Training}  \\ \hline
batch size  & 128  \\ \hline
\bf{n iters} & \bf{60000}  \\ \hline
\bf{snapshot freq} & \bf{19500}  \\ \hline
\bf{log freq} & \bf{100}  \\ \hline
eval freq & 100  \\ \hline
snapshot freq for preemption & 10000  \\ \hline
likelihood weighting & FALSE  \\ \hline
continuous training  & TRUE  \\ \hline
reduce mean & FALSE  \\ \hline
\multicolumn{2}{|c|}{Sampling}  \\ \hline
n steps each & 1  \\ \hline
noise removal & TRUE  \\ \hline
Probability flow & FALSE  \\ \hline
snr  & 0.16  \\ \hline  
\multicolumn{2}{|c|}{Evaluation}  \\ \hline
\bf{begin\_ckpt}  & \bf{1}  \\ \hline
\bf{end\_ckpt}  & \bf{1}  \\ \hline
\bf{batch size}  & \bf{128}  \\ \hline
\bf{enable sampling} & \bf{TRUE}  \\ \hline
\bf{num samples} & \bf{50048}  \\ \hline
\bf{enable loss} & \bf{FALSE}   \\ \hline
enable bpd & FALSE  \\ \hline
\multicolumn{2}{|c|}{Data}  \\ \hline
random flip & TRUE  \\ \hline
centered & FALSE  \\ \hline
uniform dequantizations & FALSE  \\ \hline
num channels & 3  \\ \hline
\multicolumn{2}{|c|}{Model}  \\ \hline
sigma min & 0.01  \\ \hline
sigma max & 50  \\ \hline
num scales & 1000  \\ \hline
beta min & \ziedit{20}{0.1} \\ \hline
beta max & 20  \\ \hline
dropout & 0.1  \\ \hline
embedding type & fourier  \\ \hline
\multicolumn{2}{|c|}{Optimization}  \\ \hline
weight decay & 0  \\ \hline
optimizer & Adam  \\ \hline
learning rate & 0.0002  \\ \hline
beta1 & 0.9  \\ \hline  
eps & 0.00000001  \\ \hline
warmup & 5000  \\ \hline
grad clip & 1  \\ \hline
\end{tabular}
\label{tab:configs1}
\end{minipage}
\begin{minipage}{0.4\textwidth}
\begin{tabular}{|l|l|}
\hline
\multicolumn{2}{|c|}{\bf{larcv\_png64\_ncsnpp\_continuous.py}} \\ \hline
\multicolumn{2}{|c|}{Training}  \\ \hline
sde & vpsde  \\ \hline
continuous & TRUE  \\ \hline
reduce  mean & TRUE  \\ \hline
\multicolumn{2}{|c|}{Sampling}  \\ \hline
method & pc  \\ \hline
predictor & euler maruyama  \\ \hline
corrector & none  \\ \hline
\multicolumn{2}{|c|}{Data}  \\ \hline
\bf{dataset} & \bf{larcv\_png64} \\ \hline
centered & TRUE  \\ \hline
\multicolumn{2}{|c|}{Model}  \\ \hline
\bf{image size} & \bf{64}  \\ \hline
name & ncsnpp  \\ \hline
scale by sigma & FALSE  \\ \hline
ema rate & 0.9999  \\ \hline
normalization & GroupNorm  \\ \hline
nonlinearity & swish  \\ \hline
nf & 128  \\ \hline
ch mult & (1, 2, 2, 2)  \\ \hline
num res blocks & 4  \\ \hline
atten resolution & (16,)  \\ \hline
resamp with conv & TRUE  \\ \hline
conditional & TRUE  \\ \hline
fir & TRUE  \\ \hline
fir kernel & [1, 3, 3, 1]  \\ \hline
skip rescale & TRUE  \\ \hline
resblock type & biggan  \\ \hline
progressive & none  \\ \hline
progressive input & residual  \\ \hline
progressive combine & sum  \\ \hline
attention type & ddpm  \\ \hline
embedding type & positional  \\ \hline
init scale & 0  \\ \hline
fourier scale & 16  \\ \hline
conv size  & 3  \\ \hline
\end{tabular}
\end{minipage}
\caption{List of all parameters used in our score model. Bolded values differ from the CIFAR-10 defaults. 
The number of training iterations (n iters) and frequency of checkpoint saving (snapshot freq) should be modified to suit the desired output. 
Running in training mode with the shown values (and our 50k LArTPC image dataset) will result in 150 epochs of training with checkpoints for generation saved every 50 epochs. 
Generating will then produce 50048 images from the first checkpoint. 
Our config files along with our code can be found on our GitHub.
}
\label{tab:configs}
\end{table*}

\begin{table*}[h]
\begin{tabular}{|c|c|c|c|c|c|c|c|c|c|}
\hline
Epochs     & 10     & 20    & 30   & 40   & 50   & 60   & 100  & 150  & 300  \\ \hline
Training   & 121.63 & 18.96 & 10.2 & 9.15 & 8.94 & 8.99 & 8.95 & 8.75 & 8.79 \\ \hline
Validation & 121.38 & 18.31 & 10.2 & 9.21 & 9.09 & 8.87 & 8.94 & 8.71 & 8.84 \\ \hline
\end{tabular}
\caption{SSNet-FID Values for all generated epochs.}
\label{appendix:SSNet-FID_values}
\end{table*}

\end{document}